\documentclass[%
 aip,
 amsmath,amssymb,
 reprint,%
]{revtex4-1}
\usepackage{graphicx}
\usepackage{dcolumn}
\usepackage{bm}
\usepackage[utf8]{inputenc}
\usepackage[T1]{fontenc}
\usepackage{mathptmx}
\usepackage{hyperref}

\usepackage{gensymb}
\usepackage{amsmath}
\usepackage{breqn}
\usepackage{float}
\usepackage[section]{placeins}

\begin{document}

\preprint{AIP/123-QED}

\title{Second Generation Readout For Large Format Photon Counting Microwave Kinetic Inductance Detectors}

\author{Neelay Fruitwala}
\email{neelay@ucsb.edu}
\affiliation{Department of Physics; University of California; Santa Barbara; California 93106; USA}
\author{Paschal Strader}
\affiliation{Dominican School of Philosophy and Theology; 2301 Vine St; Berkeley; CA 94708}
\author{Gustavo Cancelo}
\author{Ted Zmuda}
\author{Ken Treptow}
\author{Neal Wilcer}
\author{Chris Stoughton}
\affiliation{Fermi National Accelerator Laboratory; Batavia; IL 60510; USA}
\author{Alex B. Walter}
\affiliation{Jet Propulsion Laboratory; California Institute of Technology; Pasadena; California 91125; USA}
\author{Nicholas Zobrist}
\author{Giulia Collura}
\author{Isabel Lipartito}
\author{John I. Bailey III}
\author{Benjamin A. Mazin}
\affiliation{Department of Physics; University of California; Santa Barbara; California 93106; USA}

\date{\today}

\begin{abstract}
We present the development of a second generation digital readout system for photon counting microwave kinetic inductance detector (MKID) arrays operating in the optical and near-IR wavelength bands. Our system retains much of the core signal processing architecture from the first generation system, but with a significantly higher bandwidth, enabling readout of kilopixel MKID arrays. Each set of readout boards is capable of reading out 1024 MKID pixels multiplexed over 2 GHz of bandwidth; two such units can be placed in parallel to read out a full 2048 pixel microwave feedline over a 4 -- 8 GHz band. As in the first generation readout, our system is capable of identifying, analyzing, and recording photon detection events in real time with a time resolution of order a few microseconds. Here, we describe the hardware and firmware, and present an analysis of the noise properties of the system. We also present a novel algorithm for efficiently suppressing IQ mixer sidebands to below $-30\ dBc$.
\end{abstract}

\maketitle

\section{\label{sec:level1}Introduction\protect}

Microwave Kinetic Inductance Detectors (MKIDs) are a low temperature superconducting detector technology used for astronomical observations in the submilimeter through visible bands. In the optical/IR, MKIDs are single photon sensitive, energy resolving, and allow photon arrival time determination on microsecond timescales \cite{paul2017}. This makes MKIDs ideally suited to the study of highly time-variable phenomena such as pulsars \cite{mattpulsar} and compact binaries \cite{paulbinary}. MKIDs are also suited to serve as IFUs (integral field units) for exoplanet direct imaging, which requires high sensitivity (due to the extremely low flux from companions), and benefits from high time resolution (to resolve atmospheric effects). Three optical/IR MKID instruments have been built and commissioned: ARCONS (Array Camera for Optical to Near-IR Spectrophotometry, a 2 kilopixel IFU at Palomar Observatory) \cite{arcons}, DARKNESS (DARK-speckle Near-infrared Energy-resolving Superconducting Spectrophotometer, a 10 kilopixel IFU at Palomar)\cite{darkness}, and MEC (MKID Exoplanet Camera, a 20 kilopixel IFU at Subaru Observatory)\cite{mec}. 

One of the primary advantages of MKIDs over other superconducting detector technologies is the intrinsic frequency domain multiplexibility; each MKID pixel is a microwave frequency resonator with a resonant frequency that can be easily tuned in fabrication. This allows large numbers of resonators, each tuned to a different frequency, to be placed in parallel on a single microwave channel. This greatly simplifies the cryogenic wiring and electronics, shifting the complexity to the room temperature readout system \cite{benmkidog, arcons, darkness, mec}.

A first generation readout for optical MKIDs was developed in 2011 for the ARCONS instrument. This system is capable of reading out 256 channels in a 512 MHz band on each readout board \cite{seangen1}.  Eight boards were used simultaneously to readout the 2048 pixel ARCONS instrument. In order to feasibly scale up to the kilopixel arrays used by the second generation of optical MKID instruments, we have developed a new readout system capable of reading out up to 1024 channels per unit over 2 GHz of bandwidth. This system largely retains the signal processing algorithm chain found in the first generation readout, albeit with major changes to the hardware and firmware to accommodate the higher bandwidth.

\section{\label{sec:level1}Overall Approach and Requirements\protect}

An MKID is a cryogenically cooled superconducting resonator consisting of an inductor/capacitor pair. The circuit gets its inductance primarily from the kinetic energy of the Cooper pairs in the supercurrent (hence the term kinetic inductance). When a photon is absorbed by the inductor, some of the Cooper pairs are broken (creating free electrons known as ``quasiparticles"), changing the inductance and hence the resonant frequency and quality factor of the circuit. The number of quasiparticles created is proportional to the energy of the incident photon, which gives the MKID intrinsic energy resolution. To read out the detector, the resonator is driven by a microwave probe signal tuned to its resonant frequency and the probe signal is monitored for changes in its relative phase. A photon detection event manifests as a pulse in phase with a steep rise ($\sim 1\ \mu s$) and an (approximately) exponential decay with ($\tau \approx 15\ \mu s$). Our devices are designed to operate in the linear regime, where $\delta \phi \propto \delta f \propto \delta E$ \cite{day03, paul2017}. 

The full kilopixel readout generalizes the above approach to an array of resonators placed in parallel on a single microwave feedline. A comb of probe tones, one at each resonator frequency, is used to drive the MKIDs. Each resonator acts as a notch filter centered on its resonant frequency, modifying the phase and amplitude of its probe tone while leaving other tones comparatively untouched. After transmission through the array, the tones are downconverted and digitized, and each tone is independently montitored for changes in its relative phase.

The second generation of MKID arrays are in the 10-20 kilopixel range, with pixels split up into several microwave feedlines, each containing 2000 pixels inside the 4-8 GHz band. We expect a maximum photon event rate of 2500 counts/pixel/second. Due to data transmission and storage constraints, we require that each event be identified, analyzed, and recorded in real time. Following these constraints, we define the following system requirements:
\begin{itemize}
    \item Generation of a 2000 tone comb in a 4-8 GHz band, with arbitrary tone frequencies and powers.
    \item Channelization: the ability to isolate a 200-500 kHz wide channel centered around each tone
    \item Low noise: the noise floor of each channel must be lower than the device noise and cryogenic amplifier noise
    \item Sample each channel at a rate of $\sim 1\ \mu s$
    \item Implement a filtering and triggering system that is capable of detecting photon events in real time and accurately determining the associated phase pulse amplitude
\end{itemize}


\section{\label{sec:level1}System Overview\protect}

The full readout system consists of several readout units, each of which can read out up to 1024 resonator channels in a 2 GHz band. Two such units can be placed in parallel to read out a full 4 GHz feedline. Because the resonant frequency band lies outside the range of readily available ADCs, we use an IQ modulation scheme to perform up/down conversion between the digital room temperature electronics (-1 to 1 GHz) and the resonators (4 to 6 GHz or 6 to 8 GHz). Each readout unit has three major components: 1) CASPER (Collaboration for Astronomy Signal Processing and Electronics Research) ROACH-2 (Reconfigurable Open Architecture Computing Hardware) board with a Xilinx Virtex-6 FPGA for channelization and pulse detection, 2) Custom ADC/DAC board with dual 2 GSPS Analog Devices AD9625 ADCs and AD9136 DACs and Xilinx Virtex-7 FPGA for control, and 3) RF/IF board for IQ modulation. 

The basic outline of the readout procedure (illustrated in figure 1) is as follows:
\begin{enumerate}
    \item Tone generation in [-1 GHz, 1 GHz] (IF band) using dual 2 GSPS DACs.
    \item Upconversion to RF band via IQ mixing tones with local oscillator (LO) (figure \ref{rfconversion}).
    \item Tones filtered through resonators, amplified by cryogenic HEMT (high electron mobility transistor).
    \item Output is downconverted by IQ mixer, then digitized by dual 2 GSPS ADCs.
    \item ADC stream is sent to Virtex-6 FPGA for channelization, filtering, and photon triggering.
    \item Photon events (time and pixel tagged pulse height) are streamed over ethernet to data server.
\end{enumerate}

\section{\label{sec:level1}Hardware\protect}

\begin{figure*}
\centering
\includegraphics[scale=0.6]{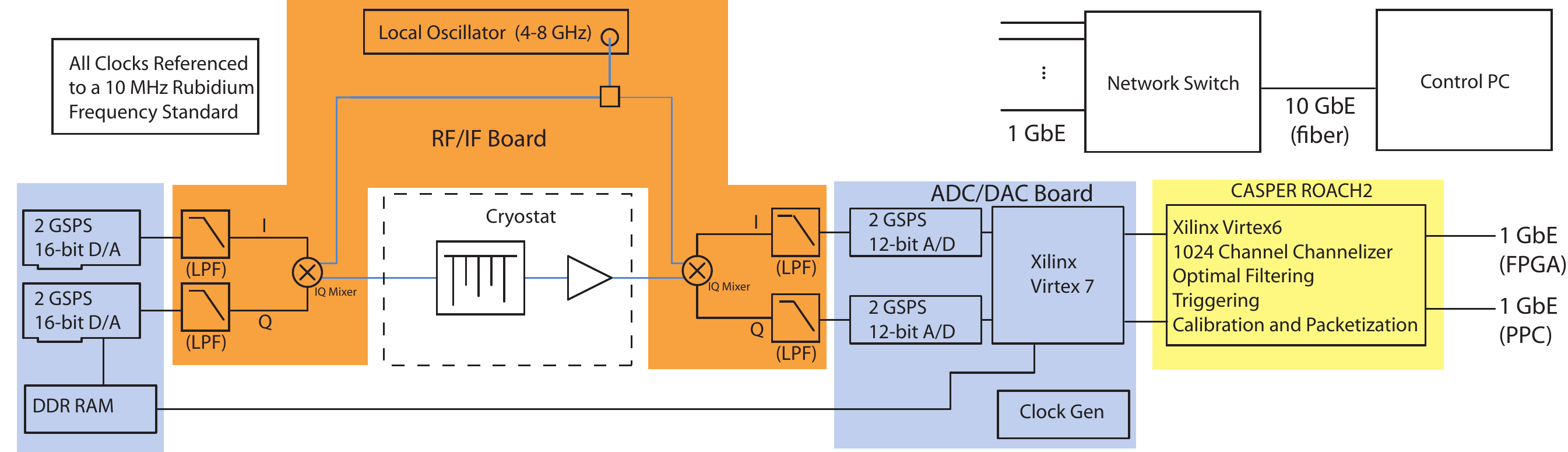}
    \caption{\label{fig:blockdiagram}System block diagram showing a single 2 GHz, 1024 channel readout unit. Reproduced with permission from  Publ. Astron. Soc. Pac 130, 988 (2018). Copyright 2018 The Astronomical Society of the Pacific.}

\end{figure*}

\subsection{\label{sec:level2}ADC/DAC Board\protect}

Due to the complexity of modern, high sample rate ADCs and DACs, we have developed a dedicated circuit board for signal routing and control of these components. The board features dual 12-bit 2 GSPS ADCs (Analog Devices AD9625), and dual 16-bit 2 GSPS DACs (Analog Devices AD9136). Dual ADC/DACs are required for complex sampling; one unit is required for each of I and Q. The board also has an onboard Xilinx Virtex-7 FPGA (XC7VX330T-2FFG1761C) for control of the ADCs and DACs, routing of the ADC signal to the ROACH-2, and communication with the ROACH-2 and RF/IF boards. A LMK04821 synthesizer is used to generate all clocks required by the ADCs, DACs, and FPGAs on both the ROACH-2 and ADC/DAC boards, synchronized to an external 10 MHz reference signal.

The DAC output is generated using lookup table (LUT) that contains the sum of all of the resonator probe tones in the time domain. The LUT is generated in software and stored in onboard DDR3 RAM. Two DACs are used to generate a complex output signal, one for I and one for Q. A table size of 262144 values (for each of I and Q) imposes a tone frequency quantization of 7.629 kHz. The LUT approach allows us to generate the DAC output for an arbitrary list of resonator frequencies, and independently specify the drive power for each resonator (required to optimize photon pulse SNR \cite{}). To generate the LUT, tones are first calculated and summed together in software according to their IF band frequencies and relative powers. Tone phases are randomized to maximize DAC dynamic range utilization. The full LUT is then scaled to fit the DAC dynamic range, and the programmable output attenuators are adjusted to set the tones to the desired power. Programmable attenuators in the RF output chain ensure full utilization of the DAC dynamic range over a wide range of tone powers. 

The IQ ADC/DAC inputs/outputs are connected to the RF/IF board via SMP connectors. There are also several SPI interfaces to enable the Virtex-7 to program the attenuators and local oscillator on the RF/IF board.

The raw ADC output is sent to the Virtex-6 FPGA on the ROACH-2 over a ZDOK connector 16 samples (8 each from I and Q) per clock in a 192-bit parallel bus clocked at 250 MHz.

\begin{figure*}[!]
\centering
\includegraphics[scale=0.5, trim={200 140 0 140}, clip]{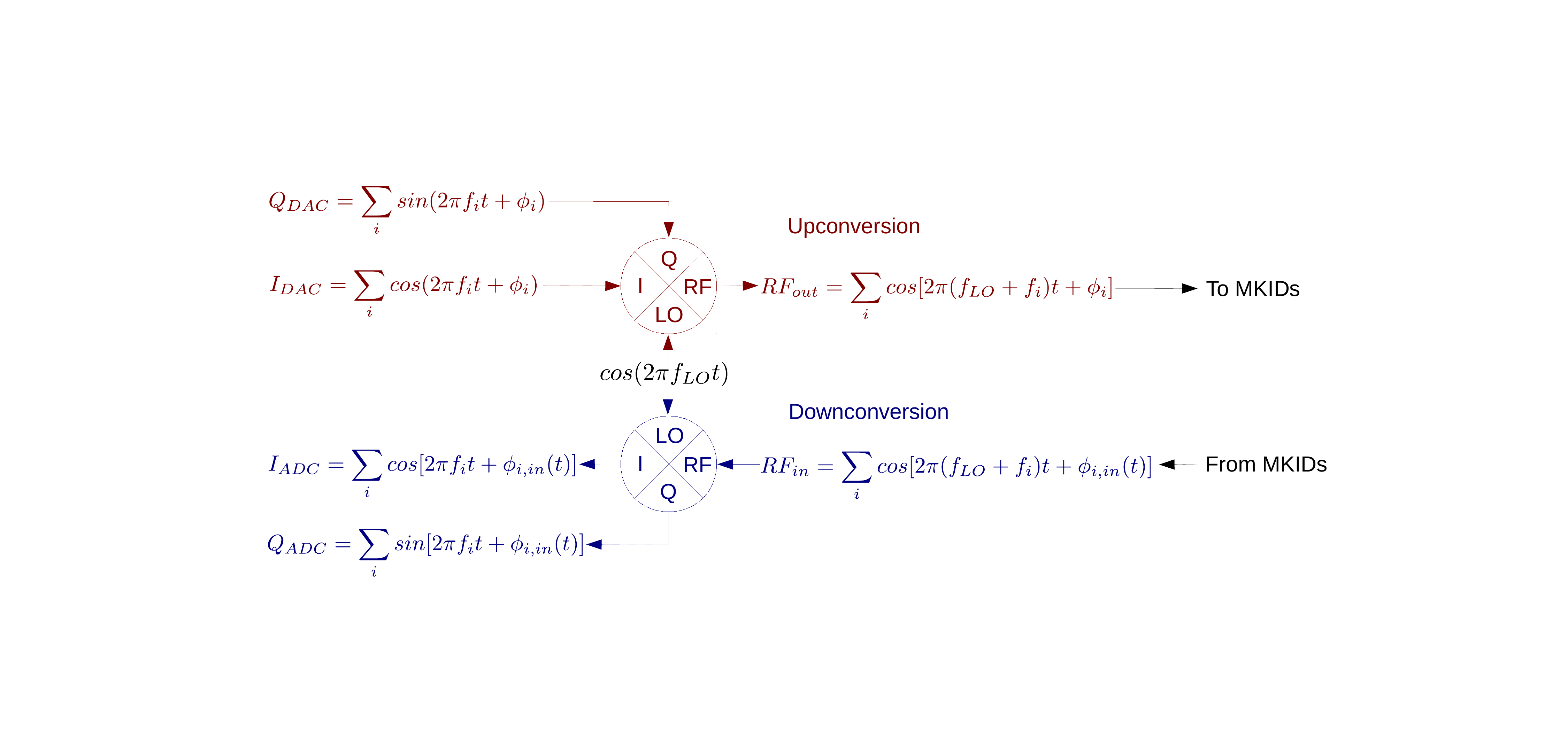}
\caption{Schematic of RF up/downconversion scheme. The frequency comb is written explicitly as a sum over the probe tones, where $f_i \in [-1\ GHz,\ 1\ GHz]$ is the IF band tone frequency, and $f_{LO} \in [4\ GHz,\ 8\ GHz]$ is the LO frequency. $\phi_i$ is the probe tone phase in the DAC LUT. $\phi_{i, in}(t) = \phi_i + \phi_{i, cable} + \delta\phi_{i,res}(t)$ is a time dependent phase term that includes the DAC LUT phase, cable delay, and phase change induced by the resonator.  The same LO is used for both upconversion and downconversion to suppress phase/frequency errors in the LO signal.}
\label{rfconversion}
\end{figure*}

\subsection{\label{sec:level2}ROACH-2 Board\protect}

The ROACH-2 was developed by the CASPER collaboration at UC Berkeley \cite{casper2011, casperbig}. It was designed around the Xilinx Virtex-6 FPGA (XC6VSX475T-1FFG1759C). It has a PowerPC CPU, DDR3 and QDR memory modules, and various communications interfaces such as ethernet. All of the core digital signal processing for the readout system (channelization, filtering, and pulse detection) is implemented on the Virtex-6 FPGA. We chose the ROACH-2 platform because CASPER provides an extensive dedicated toolset and programming interface for the FPGA. The CASPER toolflow contains blocks for performing complex signal processing tasks (such as the polyphase filter bank), as well as interfaces to much of the hardware on the ROACH-2. CASPER has also released a series of python libraries for programming and communication with the FPGA over ethernet.

\subsection{\label{sec:level2}RF/IF Board\protect}

\subsubsection{\label{sec:level3}IQ Upconversion and Downconversion\protect}

The RF/IF board implements IQ modulation for analog up/downconversion between the IF band (-1 -- 1 GHz) and the RF band (4 -- 6 GHz or 6 -- 8 GHz). For the upconversion process, the complex IF band frequency comb is mixed with a local oscillator (LO) tone at the center of the RF band (i.e. a 4 -- 6 GHz resonator band would use a 5 GHz LO), which generates a real-valued output that is shifted into the resonator frequency band. Downconversion is the reverse process, where the real valued RF output from the cryostat is mixed with the same LO to generate a complex-valued IF band signal which can be digitized by the ADCs. This process is illustrated schematically in figure 2.  The LO is generated on the board by a programmable phase locked loop (PLL) IC (Texas Instruments TRF3765). The PLL is referenced to an external 10 MHz signal provided by the ADC/DAC board. A frequency doubler is used on the output of the TRF3765 to shift the frequency range into the resonator band (600 MHz -- 9.6 GHz). Analog devices HMC525LC4 I/Q mixers are used for upconversion and downconversion. 

\subsubsection{\label{sec:level3}RF Output Chain\protect}

Two Peregrine Semiconductor PE43705 programmable step attenuators provide 0 to 63.5 dB (steps of 0.25 dB) of attenuation on the output frequency comb, which enables full utilization of DAC dynamic range over a range of resonator readout powers. The maximum output power of the full RF chain is -12.5 dBm.

\subsubsection{\label{sec:level3}RF Input Chain\protect}

In order to ensure full utilization of the ADC dynamic range over a wide range of resonator drive powers, the RF/IF board has an amplifier/attenuator chain before the IQ mixer. Four Hittite HMC3587 provide 60 dB gain, and two programmable step attenuators (63.5 dB total in steps of 0.25 dB) are used to optimize the input power at the ADC and ensure that the amplifiers are kept away from saturation (see figure \ref{rfinchain} for schematic). The optimal signal power at the end of the amplifier chain (IQ mixer input) should be approximately -7 dBm; this implies an input power range of -64 dBm to -0.5\ dBm \footnote{There is a fixed 3 dB of attenuation in addition to the programmable attenuators, so the power at the end of the chain is given by $P_{out} = P_{in} + 60\ dB - A_{prog} - 3\ dB$}. In practice, we find that in order to keep the noise from the room temperature chain sufficiently low, the total programmable attenuation should not exceed $\approx 30$ dB, providing an actual input power range of -64 dBm to -34 dBm (see Appendix A for full calculation).

\begin{figure}[!t]  
  \begin{center}
  \includegraphics[width=\columnwidth]{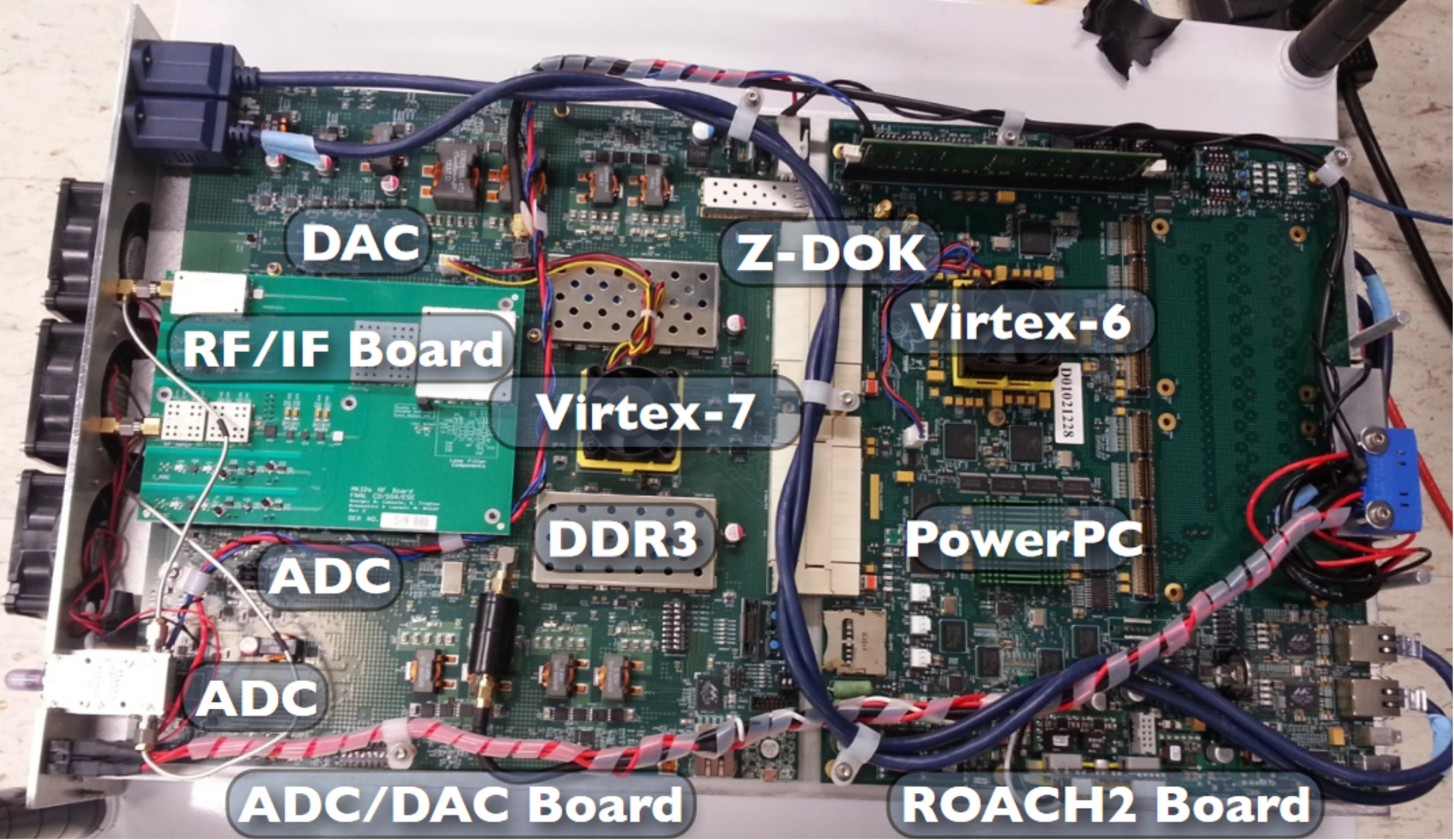}
  \end{center}
  \caption{Cartridge containing an assembled readout unit. The ROACH-2 board is connected to the ADC/DAC board by two ZDOK connectors. The RF/IF board is mounted on the ADC/DAC board using SMP blind-mate connectors for signals and GPIO pins for programming. Another readout unit is mounted to the underside of this cartridge. Figure and Caption reproduced with permission from Paschal Strader, "Digital Readout for Microwave Kinetic Inductance Detectors and Applications in High Time Resolution Astronomy", Ph.D. dissertation (University of California at Santa Barbara, 2016).}
  \label{readoutpicture}
\end{figure}

\subsection{\label{sec:level2}Timing\protect}

The ROACH-2 firmware has a timing block that keeps track of the absolute (UTC) time with microsecond precision. In order to ensure accurate absolute timing, this block can be synchronized to an external reference using a pulse-per-second (PPS) input. In our current implementation, the PPS input and the 10 MHz reference used to generate the FPGA clocks are generated by a spectracom timing module that is synchronized to a GPS satellite reference. 

\section{\label{sec:level1}Firmware and Software\protect}

\subsection{\label{sec:level2}ADC/DAC (Virtex-7 FPGA) \protect}

The Virtex-7 firmware is responsible for controlling the ADCs, DACs, and programmable components on the RF/IF board, and for routing the ADC and DAC data streams. Communication with the Virtex-7 is conducted over a UART interface to the ROACH-2 Virtex-6 FPGA, which can receive and forward commands from the control computer over ethernet. The low-level firmware and communications interfaces are controlled using a Xilinx MicroBlaze soft processor. The MicroBlaze runs a C script which initializes all major peripherals, then recieves and executes commands sent by the ROACH-2 over the UART interface. These commands include programming the RF attenuators and LO, receiving the DAC LUT and writing it to DDR3, and resetting the ADCs and DACs. The firmware was developed primarily at Fermilab, and testing and integration with the ROACH-2 were performed at both Fermilab and UCSB.

\subsection{\label{sec:level2}ROACH-2 (Virtex-6 FPGA)\protect}

The Virtex-6 firmware is responsible for the bulk of the digital signal processing performed by the system: conversion from the raw 2 GHz I/Q stream from the ADCs to time and phase (or energy) tagged photons. This process can be divided into four stages: channelization, phase conversion, optimal filtering, and photon event triggering. The firmware was developed at UCSB using the MATLAB Simulink and ISE based CASPER ROACH toolflow. All of the firmware blocks are clocked at 250 MHz using a clock supplied by the ADC/DAC board over the ZDOK connectors.

\begin{figure*}
\centering
\includegraphics[scale=0.3, trim={0, 15, 0, 35}, clip]{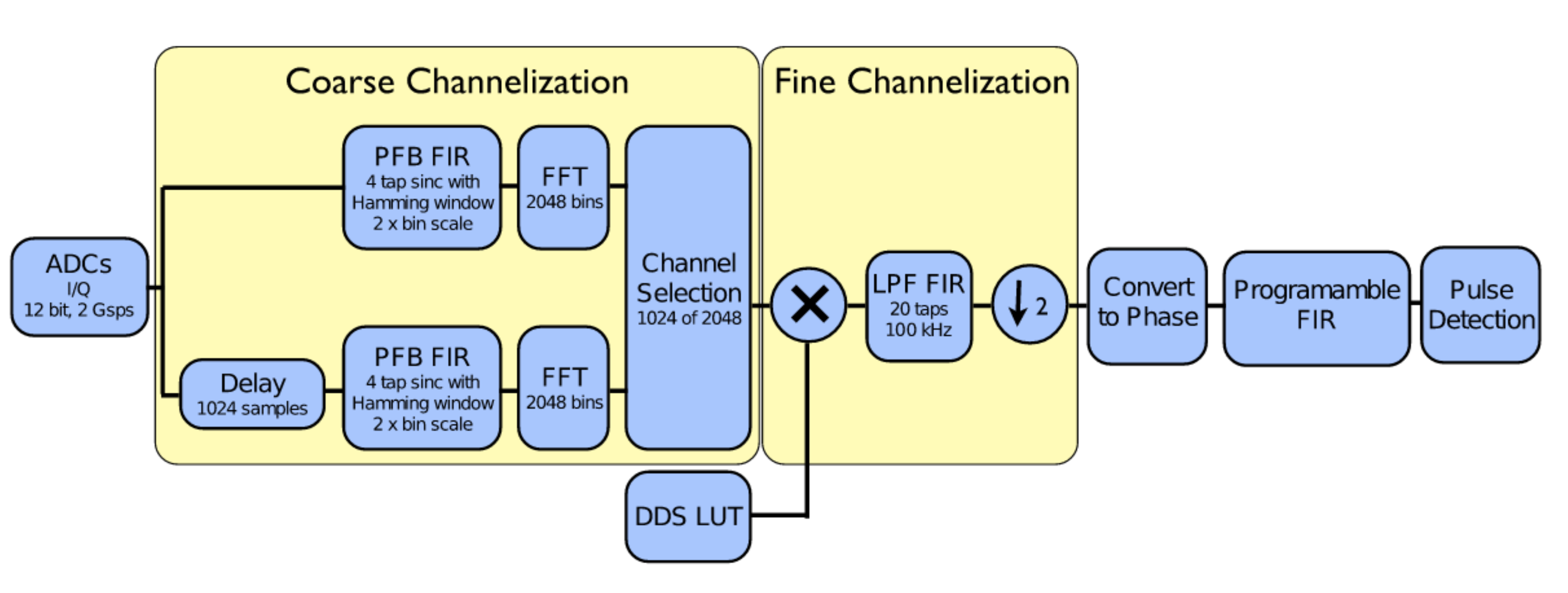}
\caption{ROACH-2 firmware block diagram. Data is streamed over the ZDOK over a 16-sample bus (8 12-bit I/Q pairs) clocked at 250 MHz. The data is then copied and processed by two parallel PFBs; one copy is delayed by 128 clocks (half of the complete FFT cycle) to maintain a uniform sampling interval of each bin at 2 MHz \cite{seangen1}. After the PFB stage, up to 1024 resonators are selected from the bins. Following channel selection, DDC is performed on each channel to center the resonator tone within the frequency bin. Channels are then low pass filtered (100 kHz on each of I and Q) and converted to phase. The phase is filtered using a channel-specific ``optimal" FIR. Phase pulses meeting the trigger conditions are recorded and streamed to the data server over 1 GBit ethernet. Figure adapted with permission from Paschal Strader, "Digital Readout for Microwave Kinetic Inductance Detectors and Applications in High Time Resolution Astronomy", Ph.D. dissertation (University of California at Santa Barbara, 2016).}

\end{figure*}

\subsubsection{\label{sec:level3}Channelization\protect}

Channelization is the process of converting the raw 2 GHz I/Q stream from the ADC into a time-multiplexed series of frequency bins, each centered around a resonator probe tone. We use the same two stage approach as the first generation readout, as described in (Ref. \onlinecite{seangen1}). The first stage is a dual 2048-sample (more specifically, 2048 branch, 4 tap) polyphase filter bank (PFB), which consists of a finite impulse response (FIR) filter followed by a Fast Fourier Transform (FFT). The configuration is the exact same as in (Ref. \onlinecite{seangen1}), except for the total bandwidth/number of points (2048 samples over 2 GHz, instead of 512 samples over 512 MHz). As in (Ref. \onlinecite{seangen1}), oversized 2 MHz bins are used, so two parallel PFBs are required to Nyquist sample each bin. Oversized bins ensure that there exists a bin with at least 500 kHz of bandwidth around every resonator, no matter its resonant frequency. The output of the PFB+FFT is time multiplexed at $(2 \times FFT)\times(8\ bins/clock) = 16\ bins/clock$. The FPGA is clocked at $250\ \textrm{MHz}$, so the sample rate of each bin is $(250\ clocks/s)\times(16\ bins/clock)/(2048\ bins) \approx 2\ \textrm{MHz}$. 

After the FFT, the bins are sorted into 1024 resonator channels; for each resonator frequency, the bin with the closest center frequency is selected. The allowed spacing between resonators can be as low as 200 kHz, so there may be as many as 10 resonators per bin. These channels are split into $4 \times 256$-channel time-multiplexed ``streams", with each resonator getting 2 samples per clock (one from each FFT). This maintains the 2 MHz channel sample rate: $(250\ \textrm{MHz})\times(2\ samples)/(256\ channels) \approx 2\ \textrm{MHz}$. Digital down conversion (DDC) is then performed on each channel to center it at 0 Hz: each channel is multiplied by a (complex) sinusoid with $f = f_{bin} - f_{res}$. These sinusoids are generated using a lookup table which is stored on the ROACH-2 QDR memory.

\subsubsection{\label{sec:level3}Low Pass Filtering and Phase Conversion\protect}

\begin{figure}
    \centering
    \includegraphics[scale=0.47, trim={10 30 50 63}, clip]{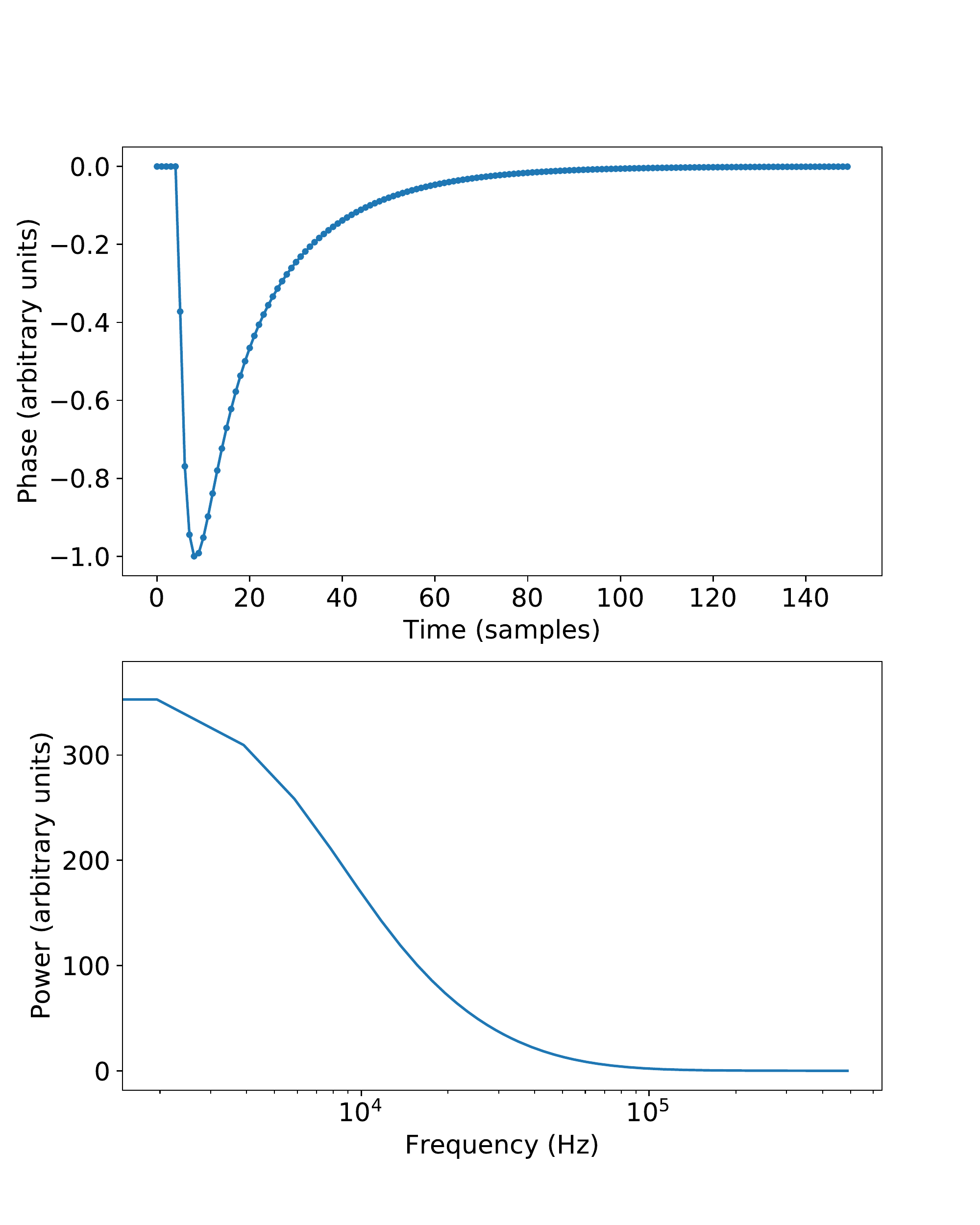}
    \caption{Top: representative phase pulse template, made by averaging together phase collected on a single resonator channel, then fitting to a triple exponential function. Bottom: power spectrum of the above pulse. Note that there is very little power above 100 kHz. }
    \label{tempfftplot}
\end{figure}

After the channelization stage, each channel is low pass filtered to set the desired channel bandwidth. This is done by separately convolving I and Q with a 20-tap Hanning-windowed sinc function. The following considerations should be taken into account when setting the channel bandwidth: 1) desired time resolution/response time of the detector (1 $\mu s$); 2) photon event (phase pulse) SNR; 3) desired minimum resonator spacing. Exoplanet imaging requires framerates of at most a few kHz ($\sim100\ \mu s$), so time resolution was not a major consideration. We found that a channel bandwidth of 200 kHz ($\pm$ 100 kHz around the tone center; which gives 100 kHz bandwidth after phase conversion) provides good performance for considerations (2) and (3). This bandwidth preserves the frequency content of a typical pulse (figure \ref{tempfftplot}) while attenuating much of the high frequency line noise present in many resonator channels (figure \ref{phasenoise}). It also allows us to impose a minimum channel spacing of 200 kHz, reducing the number of frequency collisions to $\approx 6\%$ of identified resonators. After the low pass filter, the channels are downsampled to 1 MHz, and the I/Q values are converted to phase using a CORDIC algorithm that implements $arctan(Q/I)$. 

\subsubsection{\label{sec:level3}Optimal Filtering\protect}

\begin{figure}
    \centering
    \includegraphics[scale=0.47, trim={0 0 10 51}, clip]{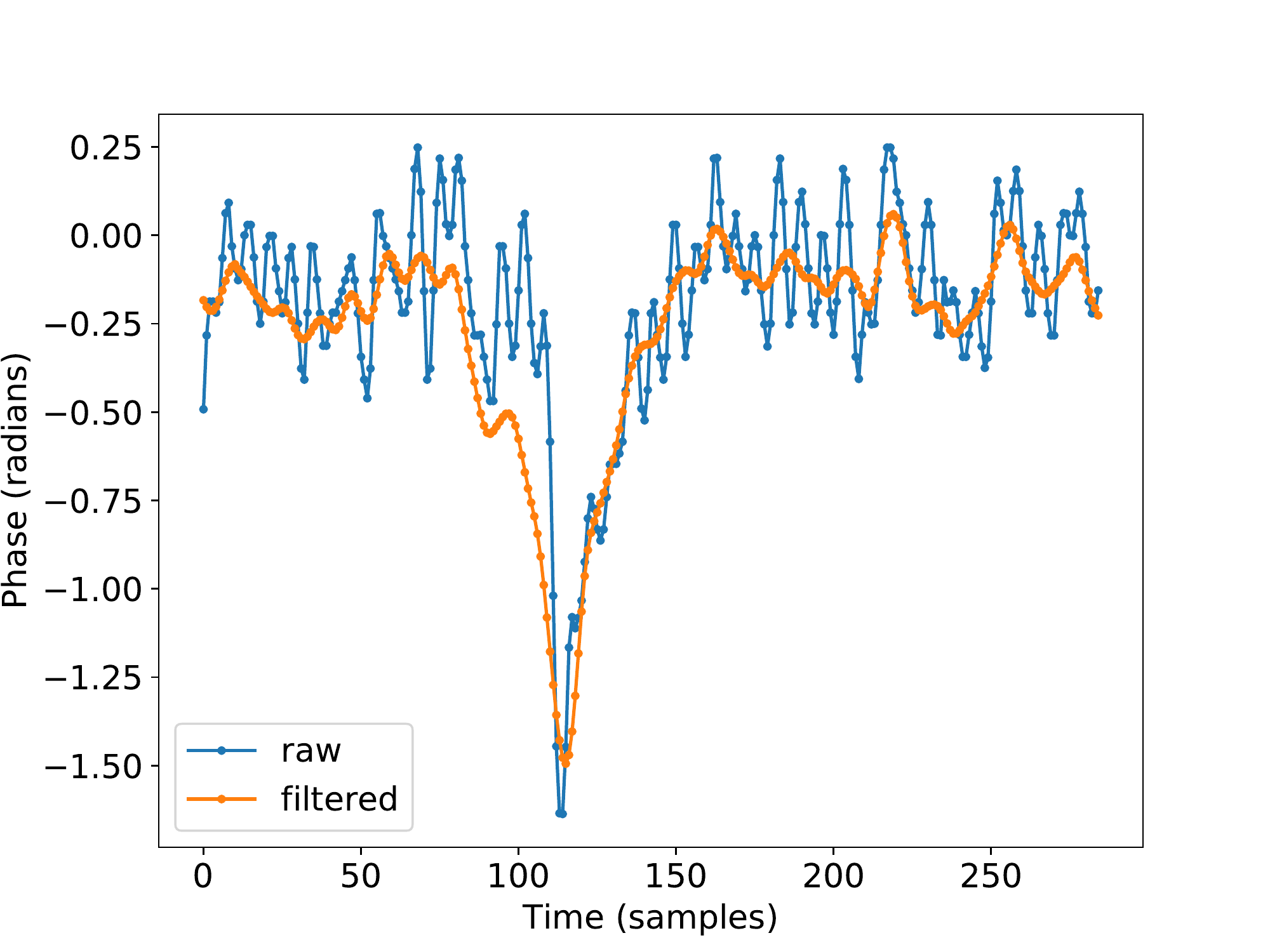}
    \caption{Example photon signal on a single resonator channel. (blue) is data collected after the 100 kHz LPF and phase conversion steps; (orange) is the same data stream after applying a 50 coefficient optimal FIR filter. Photon wavelength is 1050 nm. The sampling interval is $1.024\ \mu s$}
    \label{optfiltfig}
\end{figure}

After phase conversion, each channel is filtered using a 50-point FIR (finite impulse response) filter. The filter coefficients are programmable and are unique to each channel. This customizability is a key feature, as it allows us to tailor each filter to that channel's noise PSD and photon pulse shape. To generate these filters, we use the formalism described in (Ref. \onlinecite{origoptfilt}) and extended in (Ref. \onlinecite{alpert}). This ``optimal filter" formalism uses the noise covariance and a template of the photon pulse to produce a minimum variance linear estimator of the photon pulse amplitude. The number of filter coefficients is well matched to the characteristic pulse timescale ($\tau \approx 15\ \mu s$), as indicated by the filtered pulse being fairly symmetric (figure \ref{optfiltfig}). 

\subsubsection{\label{sec:level3}Photon Event Triggering and Streaming\protect}

The total phase data rate for a single readout unit is approximately 18 Gbps (1024 channels $\times$ 1 MHz per channel $\times$ 18 bits). This data rate is difficult to stream over ethernet (a 20 kpix array would require 360 Gbps of bandwidth) and infeasible to save for a full night of observing. To reduce the data rate, we use a firmware trigger to find and record photon pulses. The trigger consists of two conditions: 
\begin{enumerate}
    \item Local minimum detection - pulse must be preceded by 9/10 points w/ negative derivative, and immediately proceeded by two points w/ positive derivative. 
    \item Threshold - pulse amplitude must exceed a threshold value. This threshold is programmable and can be specified individually for each channel. Typically, we use some multiple of the standard deviation of the phase value $(3.5 - 4 \sigma)$.
\end{enumerate}
If the trigger conditions are met, the pulse amplitude, arrival time, and channel ID are recorded in a 64-bit data packet. Multiple such data packets are packed into UDP frames (up to 100 packets per frame) and streamed to the data server over ethernet.

\section{\label{sec:level1}Performance\protect}

\subsection{\label{sec:level1}Phase Noise\protect}

A key system performance metric is the phase noise in each resonator channel (i.e. the noise PSD of the phase timestream used for photon detection). The phase noise has three major sources: 1) noise intrinsic to the device, 2) noise contributed by the cryogenic amplification chain, 3) noise contributed by the readout system \cite{zobrist}. Ideally, phase noise is dominated by sources (1) and (2); contributions from the readout system should be negligible. To characterize this, we measure the readout system phase noise in loopback (RF input connected directly to RF output), and compare this to the expected best case performance of the device and cryogenic amplifier. 

The noise contributed by the device comes primarily in the form of a $1/f$ contribution due to two-level system (TLS) noise \cite{gaotls1, gaotls2}. We ignore this contribution, and instead use the cryogenic amplifier noise as a basis for comparison. This is because the TLS noise varies significantly between devices and is difficult to estimate, and using cryogenic amplifier noise alone provides a more stringent performance criterion than using the combination of TLS noise and cryogenic amplifier noise. For the cryogenic amplifier we assume a 2.3 K HEMT (high electron mobility transistor), which is used in both MEC \cite{mec} and DARKNESS \cite{darkness}. We assume that it has 40 dB gain and contributes thermal white noise at the device stage with a 2.3 K noise temperature (eqn. \ref{eqn:hemt_phase_noise}). We report all phase noise measurements in dBc/Hz, where dBc is the noise power relative to the probe (carrier) tone in a single quadrature. 

The noise contributed by the digital readout electronics is dominated by room temperature thermal noise and amplifier noise in the input RF chain. The exact phase noise contribution from the room temperature stage depends on the programmable attenuator settings; we calculate that for typical values the contribution of room temperature noise floor should be at least 6 dB lower than the HEMT noise contribution (see appendix A for calculation). The measured phase noise floor (figure \ref{noisevsfreq}) is $\approx 3$ dB higher than expected, likely due to unaccounted for attenuation in the RF input chain.

\begin{figure}[!]
\centering
\includegraphics[scale=0.52, trim={0 0 5 35}, clip]{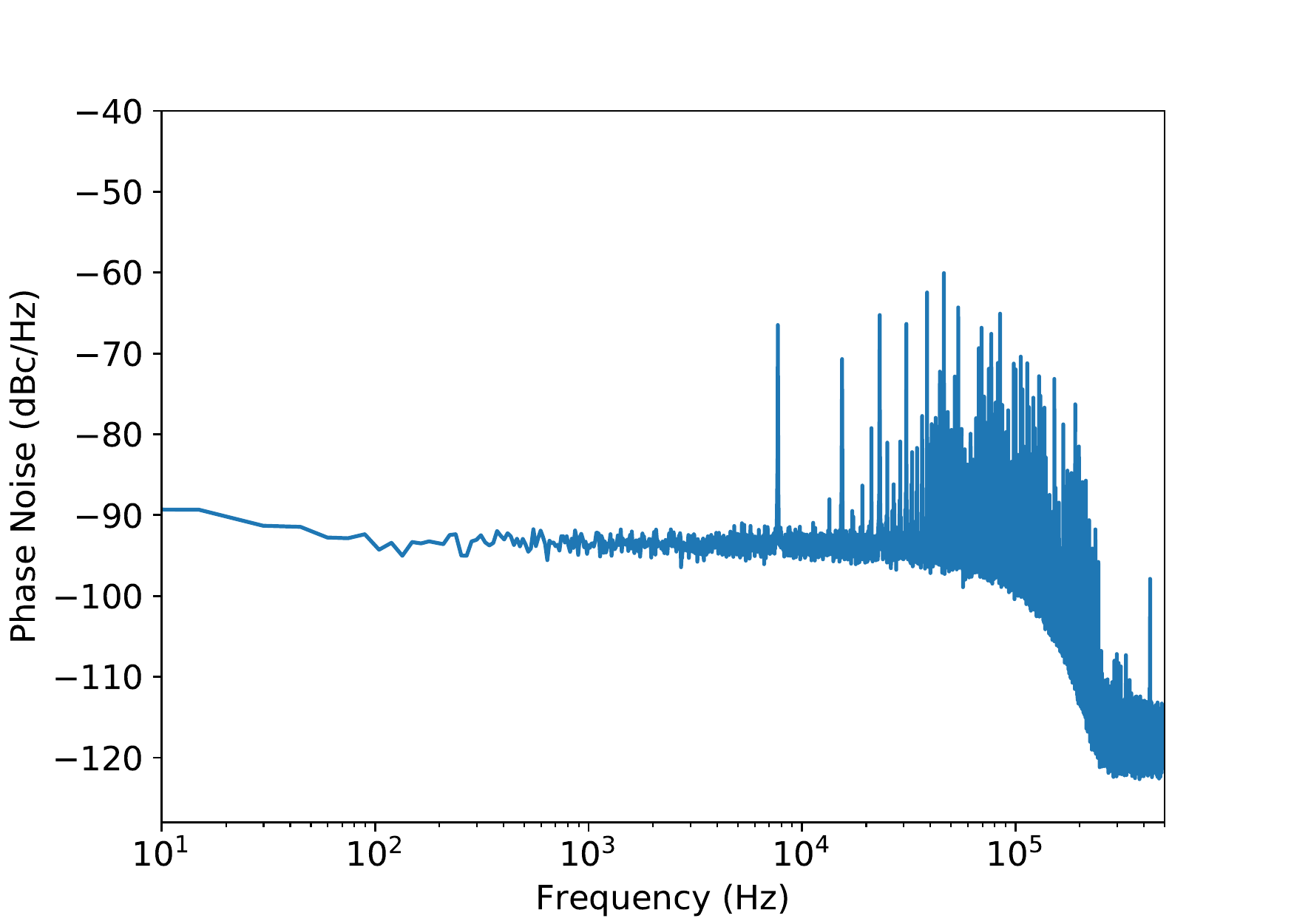}
    \caption{Phase noise spectrum of a single resonator channel. Measurement was taken in loopback with a full 1024 tone comb in $4-6\ GHz$ band. Aside from spurious signals, the noise floor is approximately -93 dBc/Hz.}
    \label{phasenoise}
\end{figure}

\begin{figure}[!]
\centering
\includegraphics[scale=0.52, trim={0 0 5 20}, clip]{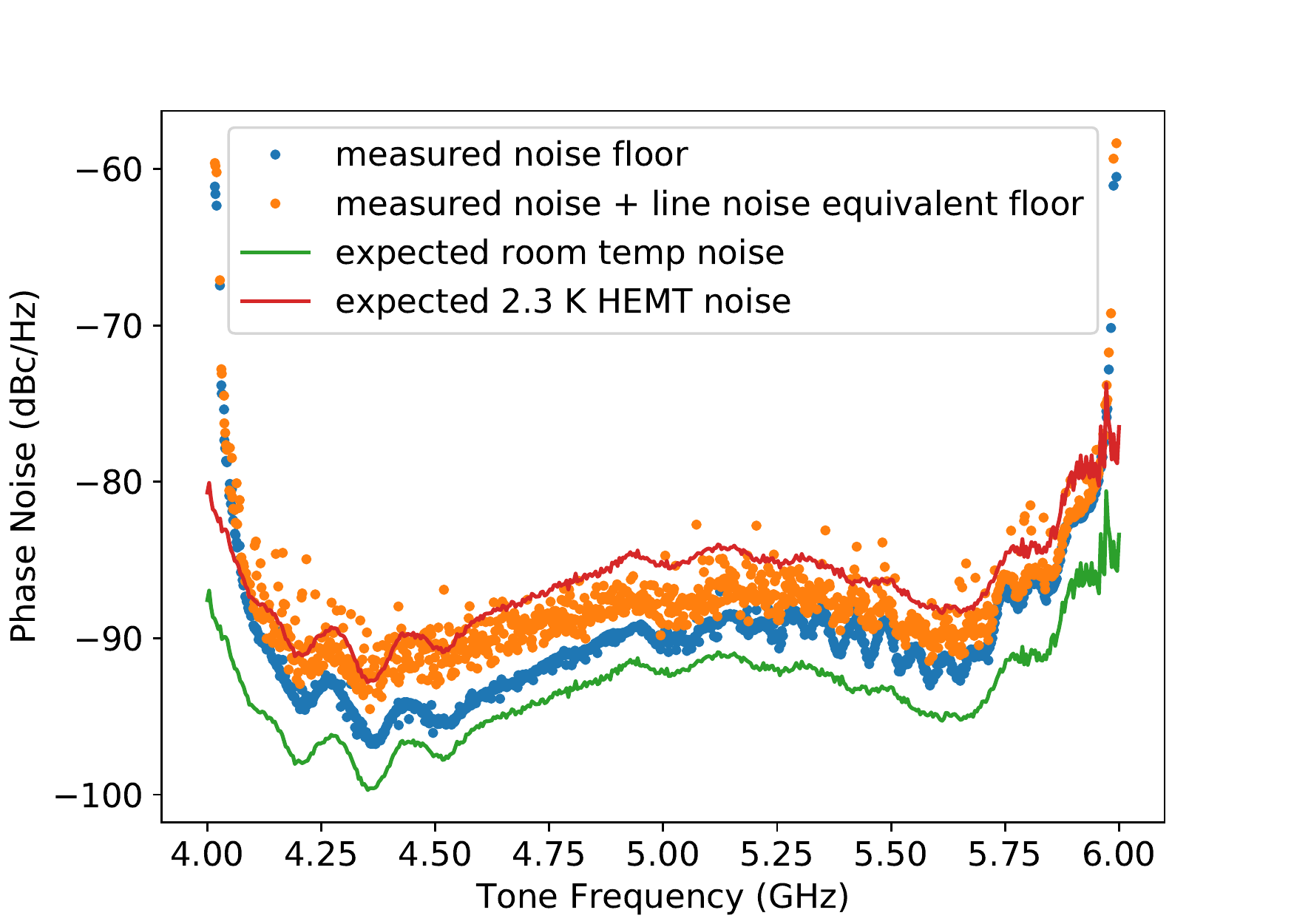}
    \caption{Loopback phase noise measurement for 1024 tone comb on a single readout unit. The measured phase noise value for each channel was computed by fitting the white noise component of the phase noise spectrum (example spectrum in figure \ref{phasenoise}). The equivalent phase noise floor after accounting for line noise is also plotted (calculation in appendix B). The expected phase noise contributions from the HEMT and room temperature chain were estimated using the measured tone power at the RF input (see appendix A for calculation).}
    \label{noisevsfreq}
\end{figure}

To verify system performance, we performed a loopback measurement of the phase noise of each tone in a full 1024 tone frequency comb. The tones were given frequencies in 4 -- 6 GHz range and uniform output power in the DAC LUT. The output power of each RF tone ranged from approximately -65 dBm to -70 dBm, where the variation results from the nonuniform frequency response of the RF output chain (which we calibrate out when reading out actual devices). The exact readout power (power at the MKID) that this corresponds to depends on the array properties and experimental setup, but the typical RF attenuation before the device is $\approx 35$ dB, which corresponds to a readout power of approximately $-100$ dBm to $-105$ dBm. Results are plotted in figure \ref{noisevsfreq}, along with expected contributions from the room temperature amplifier chain and HEMTs.

The measured phase noise has significant line noise (figure \ref{phasenoise}). The line noise occurs at multiples of 7.629 kHz, which is the quantization frequency of the DAC comb. So, we suspect that the lines are generated as intermodulation products in the input RF chain. In order to compare the line noise to the HEMT and room temperature noise, we examine the effect each has on the variance of the measured photon pulse amplitude. We do this by removing the spectral lines from the noise PSD, then scaling this ``flattened" PSD such that its corresponding pulse height estimator variance is equal to that of the original PSD (see appendix B for full calculation).

There is a significant apparent difference between our phase noise measurements ($\approx -93$ dBc/Hz), and the measured phase noise for the first generation system ($-106$ dBc/Hz, including HEMT noise \cite{seangen1}). This is because the first generation system measurements were performed at $-85$ dBm \cite{seangen1} (significantly higher than used in practice with ARCONS), and our measurements were performed at an expected readout power of $-100$ dBm to $-105$ dBm (which is close to the $-106$ dBm readout power \cite{zobrist} of the PtSi devices used in DARKNESS \cite{darkness} and MEC \cite{mec}). So, we expect that any thermal noise source (which includes both HEMT and room temperature amplifier noise) will contribute up to 20 dB higher phase noise (phase noise is inversely proportional to tone power; see eqns. \ref{eqn:hemt_phase_noise}, \ref{eqn:hemt_phase_noise_rt}, \ref{eqn:rt_phase_noise}) for readout tone powers characteristic of our system as compared to the first generation system; this is reflected in our measurements in figures \ref{phasenoise} and \ref{noisevsfreq}.

It is difficult to determine the effect of readout system noise on the ultimate resolving power of the MKIDs, as this depends on detector sensitivity (i.e. pulse height as a function of photon energy), quality factor, and device noise, which vary significantly between pixels \cite{mec, paul2017}. However, the majority of channels are below the expected HEMT noise floor, so these would obey the HEMT noise limits measured in (Ref. \onlinecite{zobrist}). We also note that the measurements in figures \ref{phasenoise} and \ref{noisevsfreq} are close to worst-case performance, with RF input attenuators $A_1 = 14.25$ dB and $A_2 = 14.5$ dB (see figure \ref{rfinchain} for block diagram of RF input chain). A more typical use case would have $A_{1,2} < 10$ dB, which will reduce the phase noise floor by $> 2.7$ dB (eqn. \ref{eqn:hemtvsrt}), making almost every channel HEMT noise limited. 

\section{\label{sec:level1}Sideband Suppression\protect}

Imperfections in IQ mixer performance lead to the presence of ``sidebands", or reflections across the LO frequency when the mixer is used to upconvert or downconvert RF tones. For example, a 6.5 GHz tone generated using a 6 GHz LO will have a sideband tone at 5.5 GHz (which may also show up at -500 MHz in the IF band after downconversion). Ideally, these sidebands are $>30$ dB down from the original tone. In our system, we have observed sidebands with  powers as high as $10 - 15$ dB below the original tone. A sideband is problematic if it happens to be within 100 kHz of another tone, as it will leak into that channel and cause its phase timestream to oscillate at $f = f_{tone} - f_{sideband}$, degrading the performance of that pixel. 

\begin{figure}
\centering
\includegraphics[scale=0.35, trim={0 20 0 50}, clip]{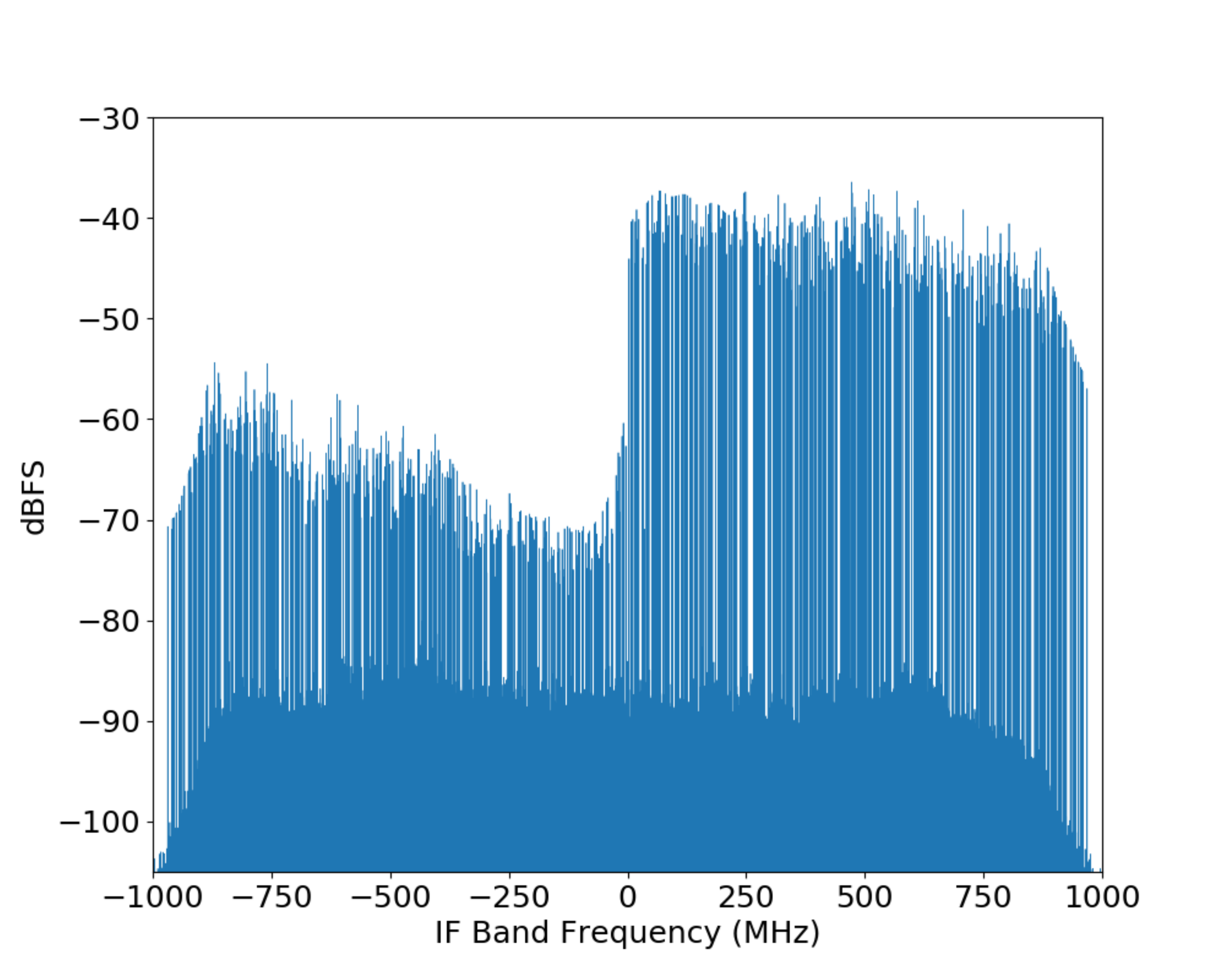}
\caption{Example sideband spectrum (taken using the ADC on loopback). Tones on the right half $(f > 0)$ are actual tones generated by the DAC, while tones on the left half are undesirable sideband reflections of these generated tones. In the worst case, sideband power is only 10-15 dB down from the tone power.}
\end{figure}

The input to the I/Q mixer can be adjusted to compensate for mixer errors and suppress sidebands. This can be done for both upconversion and downconversion:
\begin{itemize}
    \item Upconversion: adjust the relative phase/amplitude of the I and Q inputs to the mixer. This effectively introduces sidebands in the IF input to cancel them out in the output.
    \item Downconversion: introduce real RF sidebands to cancel out sidebands produced in the downconverted I and Q.
\end{itemize}
Performing these adjustments separately for upconversion and downconversion is infeasible, as injecting RF sidebands after upconversion would require major design changes. So, we instead adjust the output I and Q generated by the DAC to minimize total sideband power remaining after both upconversion and downconversion. 

For our system, we experimentally determined that the optimal phase and amplitude adjustments are sensitive to tone power and frequency and are not always smooth functions of these. So, a global fitting approach would be suboptimal, and any adjustments must be made on a per-tone basis in the DAC LUT. We have developed an algorithm to perform this optimization for an arbitrary list of tone frequencies and powers. Our algorithm minimizes total sideband power remaining after both RF upconversion and downconversion (and everything else that might be in the RF chain, including resonators, etc). The optimization problem for $N$ tones can be written as: 
\begin{equation}
\max_{\Delta\phi_{1},r_{IQ,1}... \Delta\phi_{N},r_{IQ,N}}\sum_{i} P_i - P_{sideband, i}(\Delta\phi_{i}, r_{IQ,i})
\end{equation}
where $i$ indexes the tones, $P_i$ is the tone power, and $P_{sideband, i}(\Delta\phi_{i}, r_{IQ,i})$ is the sideband power of tone $i$ as a function of the phase offset $\Delta\phi_{IQ, i}$ (from $90 \degree$) between $I_i$ and $Q_i$ and $r_{IQ,i} = |I_i/Q_i|$ ($I_i$ and $Q_i$ are the DAC outputs for tone $i$). Sideband power is measured in the input IF band by the ADCs, in units dBFS (dB relative to ADC full scale). To simplify the problem, we have assumed that the tones are independent (i.e. adjusting the phase or amplitude balance of one tone will not affect the sideband power of any other tone); so the problem can be thought of as a series of $N$ independent maximizations over the two dimensional domain given by $\{\Delta\phi_{IQ} \times r_{IQ}\}$. 

The following procedure is used to evaluate the objective function for a specific set of $\{(\Delta\phi_{IQ,1}, r_{IQ,1}), \allowbreak (\Delta\phi_{IQ,2}, r_{IQ,2}), \hdots, \allowbreak (\Delta\phi_{IQ, N}, r_{IQ,N})\}$: 
\begin{enumerate}
    \item Load DAC LUT containing these offsets
    \item Take a 8,388,608-point snapshot (set by the amount of available memory) of the ADC input
    \item Take an FFT of the snapshot to find sideband powers
\end{enumerate} 

\begin{figure}
\includegraphics[scale=0.4, trim={140 60 150 100}, clip]{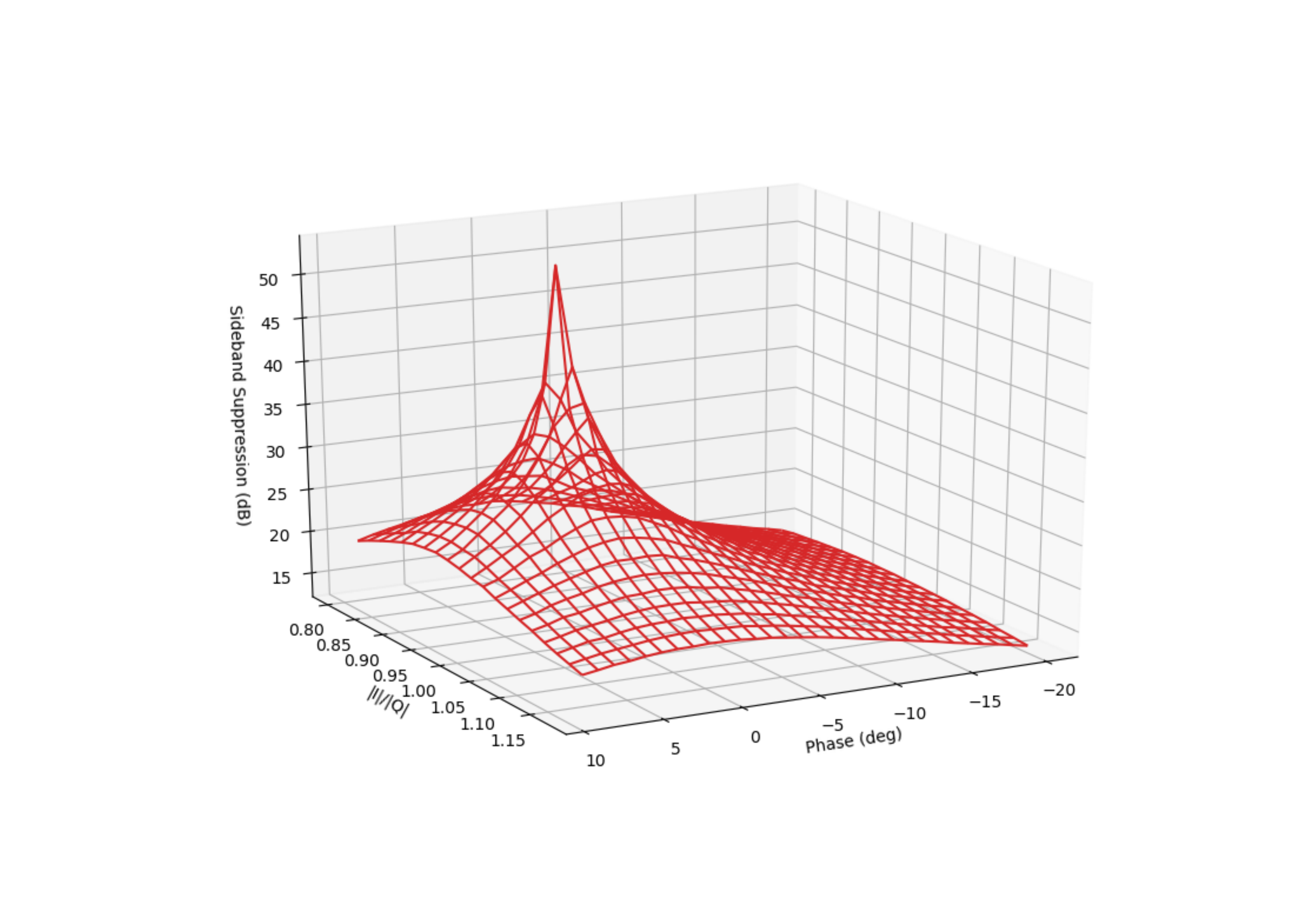}
    \caption{Plot of sideband suppression of a single tone as a function of $\phi$ and $r_{IQ}$. Close to the optimal value, the sideband suppression decays approximately exponentially, which is why an exponential fit is used in the search algorithm.}
    \label{sbgridtone}
\end{figure}

This evaluation process takes approximately one minute, so a brute force optimization with the required range and precision (phase: $[-20 \degree, 20 \degree]$, $1 \degree$ increments; amplitude ratio: $[0.8, 1.2]$ in increments of 0.2) would take approximately 13 hours. Since the optimization must be repeated for each new configuration of tones, the brute force approach is infeasible and a more efficient algorithm is needed. We have developed the following approach:

\begin{figure*}
    \centering
    \includegraphics[scale=1, trim={70 210 50 200}, clip]{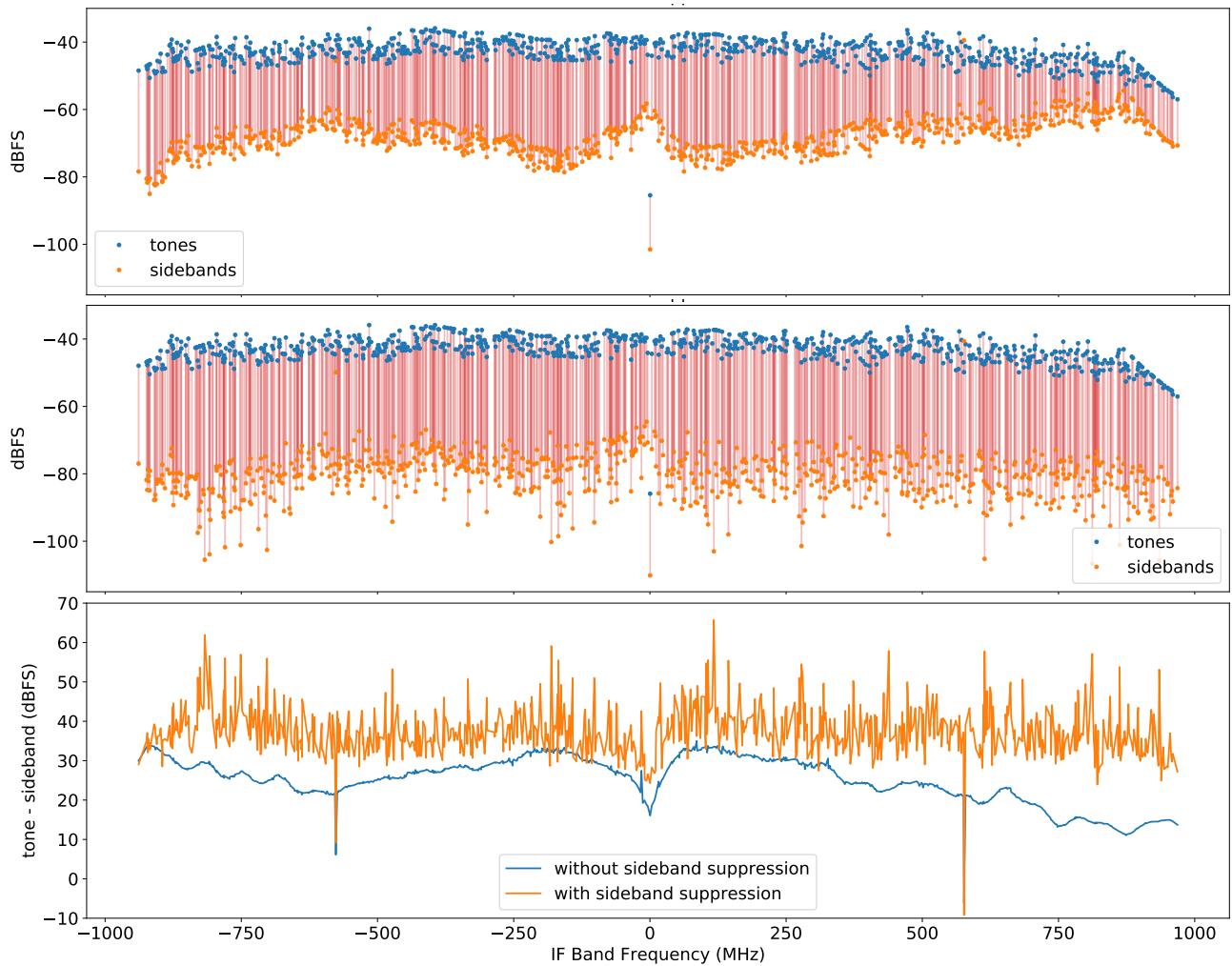}
    \caption{Test of the sideband suppression algorithm on loopback using a 868 tone MEC array frequency comb. Top/middle: comparison of each tone's power with its corresponding sidedband power, before (top) and after (middle) running the sideband suppression algorithm. Bottom: Magnitude of sideband suppression across the frequency comb.}
    \label{sbsupresult}
\end{figure*}

\begin{enumerate}
    \item Sample the objective function at $n$ randomly chosen sets of $\{(\Delta\phi_{IQ,1}, r_{IQ,1}), \allowbreak (\Delta\phi_{IQ,2}, r_{IQ,2}), \hdots, \allowbreak (\Delta\phi_{IQ, N}, r_{IQ,N})\}$ (in addition to $(\Delta\phi_{IQ,i}, r_{IQ,i}) = (0, 0)$ for all $i$)
    \item For each tone, fit $P_i - P_{sideband, i}(\Delta\phi_{i}, r_{IQ,i})$ to an exponential decay function (see figure \ref{sbgridtone}).
    \item Randomly sample a new set of $\{(\Delta\phi_{IQ,1}, r_{IQ,1}), \allowbreak (\Delta\phi_{IQ,2}, r_{IQ,2}), \hdots, \allowbreak (\Delta\phi_{IQ, N}, r_{IQ,N})\}$, either close to the fit centers or uniformly over the domain (choose this with some probability $\epsilon$).
    \item Repeat (2) and (3) until all sidebands are sufficiently suppressed (typically $P_{tone} - P_{sideband} > 30$ dB), or the number of iterations has reached some threshold value.
\end{enumerate}

Using this approach, we are able to suppress sidebands by at least 30 dB for almost every tone, with a convergence time of 30 - 45 minutes. Results from a loopback test using a MEC array frequency comb with 868 tones are shown in figure \ref{sbsupresult}. It is difficult to provide a general estimate of the improvements to energy resolution resulting from suppressing sidebands to $\geq 30$ dBc, as this depends on highly variable device characteristics (including sensitivity, quality factor, device noise), as well as the location of of the interfering sideband in the spectrum of a nearby resonator channel. We feel that the 30 dBc benchmark is appropriate, as it has an equivalent RMS phase $> 3$ dB below that of a typical resonator channel (appendix C), and is the same benchmark used for the first generation system \cite{seangen1}.

\section{Conclusion}
We have developed a digital readout system which, for the first time, makes it feasible to read out kilopixel-scale photon counting optical/IR MKID arrays. Our system is actively being used by two MKID cameras; MEC at Subaru Observatory, and DARKNESS at Palomar Observatory. All of our array calibration and control software, as well as the Virtex-6 firmware, is open source and can be found at: \url{https://github.com/MazinLab/MKIDReadout}. 

\section{Acknowledgements}

The authors would like to thank Jonas Zmuidzinas for valuable feedback regarding the phase noise calculations. NF is supported by a grant from the Heising-Simons Foundation. NZ was supported throughout this work by a NASA Space Technology Research Fellowship. IL was supported by the National Science Foundation Graduate Research Fellowship under grant number 1650114. This work was funded in part by an NSF ATI grant AST-1308556, and by US Department of Energy, Office of Science, Office of High Energy Physics contract No. DE-AC02-07CH11359 w/ Fermi Research Alliance, LLC. We would also like to thank Xilinx for donating the FPGAs used for this research.

\section{Data Availability Statement}
The data that support the findings of this study are available from the corresponding author upon reasonable request.

\appendix
\section{Phase Noise Calculations}

The single-sided phase noise spectral density $S_{\delta\phi}$ due to Johnson noise from the HEMT is given by \cite{jonasreview}:

\begin{equation}
S_{\delta \phi, HEMT} = \frac{k_B}{P_{device}} (T_{device} + T_{HEMT})
\label{eqn:hemt_phase_noise}
\end{equation}

To facilitate comparison with loopback data taken at room temperature, we can instead use the tone power at the readout system RF input; $P_{RF} = G_{HEMT} P_{device}$, where $G_{HEMT}$ is the HEMT amplifier gain. So,

\begin{equation}
S_{\delta \phi, HEMT} = \frac{k_B G_{HEMT}}{P_{RF}} (T_{device} + T_{HEMT})
\label{eqn:hemt_phase_noise_rt}
\end{equation}

\begin{figure*}
\centering
\includegraphics[scale=0.4, trim={10 320 5 350}, clip]{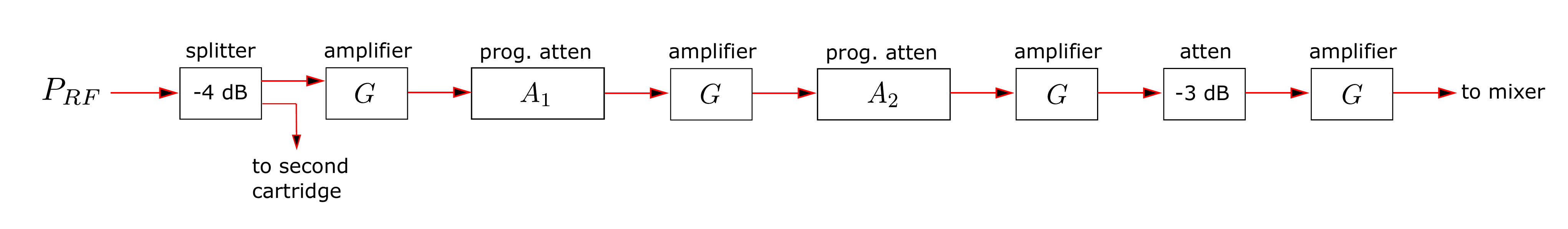}
    \caption{Schematic of room temperature amplifier/attenuator chain. An RF power splitter is used to send the full $4-8\ GHz$ output comb to both the $4-6\ GHz$ and $6-8\ GHz$ readout units. The amplifier gain $G$ is 15 dB and $0 \leq A_{1,2} \leq 31.75$ dB. In practice, we require $A_{1,2} \leq 15$ dB.} 
    \label{rfinchain}
\end{figure*}

The noise from the readout system is dominated by the room temperature amplifier chain immediately following the RF input of the RF/IF board. This chain consists of four HMC3587, each with a noise temperature of 438 K and two programmable attenuators (figure \ref{rfinchain}). The total single sided noise spectral density at the end of this chain (IQ mixer input) is given by:

\begin{dmath}
    S_{mixer} = G \left\{ \frac{G}{2} \left(\frac{G}{A_2} \left[\frac{G}{A_1} \left(S_{RT} + S_{amp} \right) + S_{RT} + S_{amp} \right] + S_{RT} + S_{amp} \right) + S_{RT} + S_{amp} \right\}
              = \left(\frac{G^4}{2 A_1 A_2} + \frac{G^3}{2 A_2} + \frac{G^2}{2} + G \right)(S_{RT} + S_{amp})
\end{dmath}

\noindent where $S_{RT, amp} = 2 k_b T_{RT, amp}$ is the Johnson noise PSD at room temperature and from the amplifier, respectively, $G$ is the room temperature amplifier gain, $A_{1,2}$ are the programmable attenuator values, and $T_{RT, amp}$ are noise temperatures at room temperature (290 K) and of the amplifiers (438 K). 

The tone power at the end of the amplifier chain is given by: 
\begin{equation}
P_{mixer} = \frac{G^4}{5 A_1 A_2} P_{RF}
\end{equation}

To obtain the single quadrature phase noise spectral density, we have:

\begin{dmath}
S_{\delta\phi, RT} = \frac{S_{mixer}}{2 P_{mixer}} 
                   = \frac{k_B}{P_{RF}}\left( \frac{5}{2} + \frac{5 A_1}{2 G} + \frac{5 A_1 A_2}{2 G^2} + \frac{5 A_1 A_2}{G^3} \right) (T_{RT} + T_{amp})
\label{eqn:rt_phase_noise}
\end{dmath}

Comparing the HEMT and room temperature terms:

\begin{dmath}
    \label{eqn:hemtvsrt}
    \frac{S_{\delta\phi, RT}}{S_{\delta\phi, HEMT}} = \left( \frac{5}{2} + \frac{5 A_1}{2 G} + \frac{5 A_1 A_2}{2 G^2} + \frac{5 A_1 A_2}{G^3} \right)\frac{T_{RT} + T_{amp}}{G_{HEMT}(T_{device} + T_{HEMT})}
\end{dmath}

In normal operation we require $A_{1,2} \leq G$. So we can set $A_{1,2} = G$ to determine an upper bound on $S_{\delta\phi, RT}$:

\begin{dmath}
    \frac{S_{\delta\phi, RT}}{S_{\delta\phi, HEMT}} \leq \left(\frac{15}{2} + \frac{5}{G} \right)\frac{T_{RT} + T_{amp}}{G_{HEMT}(T_{device} + T_{HEMT})}
\end{dmath}

For $T_{RT} = 290\ K$, $T_{amp} = 438\ K$, $T_{device} = 0.1\ K$, $T_{HEMT} = 2.3\ K$, $G_{HEMT} = 40\ dB$, and $G = 15\ dB$, $\frac{S_{\delta\phi, RT}}{S_{\delta\phi, HEMT}} \leq -6.3\ dB$.

\section{Correcting for Line Noise}

According to the optimal filtering formalism, the pulse amplitude estimator variance is given by \cite{golwalathesis}:
\begin{equation}
    \sigma^2 = \left(\sum_{k}{\frac{|\hat{s}(f_k)|^2|}{J(f_k)}}\Delta f\right)^{-1}
    \label{varest}
\end{equation}
\noindent where $k$ indexes frequency, $\Delta f$ is the frequency spacing, $\hat{s}(f_k)$ is the DFT (discrete Fourier transform) of the pulse template, and $J(f_k)$ is the noise PSD.

Consider some $J(f_k)$ with significant spectral lines. We can remove these lines -- denote this $J'(f_k)$ -- then compute a new pulse height variance $\sigma^2[J'(f_k)]$, where we've written $\sigma^2$ as a function of $J(f_k)$. Define:
\begin{equation}
    C = \frac{\sigma^2[J(f_k)]}{\sigma^2[J'(f_k)]}
\end{equation}
Since $\sigma^2$ scales proportionally with $J(f_k)$,
\begin{equation}
    \sigma^2[J(f_k)] = C \sigma^2[J'(f_k)] = \sigma^2[C J'(f_k)]
\end{equation}

\noindent $CJ'(f_k)$ is a ``clean" spectrum with the same $\sigma^2$ as $J(f_k)$, so we define the noise floor of $CJ'(f_k)$ as the ``true" channel noise floor after accounting for spectral lines. Since $J'(f_k)$ has the same noise floor as $J(f_k)$, we can perform this correction by scaling the fitted noise floor of $J(f_k)$ by $C$. 

To compute $J'(f_k)$, we set all spectral lines that are $3\ dB$ above the fitted noise floor to the local spectral floor. This correction is performed over $500\ Hz$ to $300\ kHz$ band; it does not account for the $1/f$ ``knee" present in some channels. The IQ low pass filter rolls off transmission significantly past $300\ kHz$, so it is not necessary to remove lines in this region.

In order to keep this calculation consistent with the optimal filtering scheme used in firmware, we calculate $\sigma^2$ directly from the optimal filter and pulse template. In the frequency domain, the optimal filter is given by:
\begin{equation}
    \hat{\phi}(f_k) = \frac{\hat{s}^*(f_k)}{J(f_k)}
    \label{optfilt}
\end{equation}
Substituting eqn. \ref{optfilt} into \ref{varest}, we have:
\begin{equation}
    \sigma^2 = \left(\sum_{k}{\hat{\phi}(f_k)\hat{s}(f_k)\Delta f}\right)^{-1}
    \label{varestfilt}
\end{equation}

We can convert this to the time domain using the convolution theorem (up to a normalization constant, since we are only interested in the ratio $\sigma^2[J(f_k)]/\sigma^2[J'(f_k)]$):
\begin{dmath}
    \sigma^2 = \left.\left(\sum_{k}{\hat{\phi}(f_k)\hat{s}(f_k) e^{i f_k t}\Delta f}\right)^{-1}\right\vert_{t=0} 
    = \left.iFFT[\hat{\phi}(f_k)\hat{s}(f_k)](t)\right\vert_{t=0}
    = \left.[\phi*s](t)\right\vert_{t=0}
    \label{varconv}
\end{dmath}
\noindent where $*$ denotes the convolution operator. We use equation \ref{varconv} to compute all $\sigma^2$, using a 50 coefficient optimal filter and representative pulse template (figure \ref{tempfftplot}). To avoid potential periodicity artifacts inherent to DFT, optimal filters are computed in the time domain \cite{alpert}.


\section{Noise Contributed by Sideband Tones}

Consider a sideband reflection falling within $100\ kHz$ (inside the channel bandwidth) of some resonator tone. The power contributed by the sideband reflection to the resonator tone phase power spectrum (in units dBc) is given by:
\begin{equation}
    P_{SB, \delta\phi} = \frac{1}{2}\frac{P_{SB}}{P_{tone}}
\end{equation}
where $P_{tone}$ resonator tone power and $P_{SB}$ is the sideband reflection power. The corresponding RMS amplitude (in radians) is given by:
\begin{equation}
    A_{SB, \delta\phi} = \sqrt{P_{SB, \delta\phi}} = \sqrt{\frac{P_{SB}}{2 P_{tone}}}
\end{equation}

For $\frac{P_{SB}}{P_{tone}} = -30\ dB$, $A_{SB, \delta\phi} = 1.3 \degree$. This is below the RMS phase noise for a typical resonator channel; the channel in figure \ref{phasenoise} has $A_{\delta\phi, RMS} \approx 2 \degree$. Depending on the location of the sideband reflection relative to the in-band tone, this can likely be further reduced by optimal filtering.

\nocite{*}
\bibliography{references}

\begin{thebibliography}{20}%
\makeatletter
\providecommand \@ifxundefined [1]{%
 \@ifx{#1\undefined}
}%
\providecommand \@ifnum [1]{%
 \ifnum #1\expandafter \@firstoftwo
 \else \expandafter \@secondoftwo
 \fi
}%
\providecommand \@ifx [1]{%
 \ifx #1\expandafter \@firstoftwo
 \else \expandafter \@secondoftwo
 \fi
}%
\providecommand \natexlab [1]{#1}%
\providecommand \enquote  [1]{``#1''}%
\providecommand \bibnamefont  [1]{#1}%
\providecommand \bibfnamefont [1]{#1}%
\providecommand \citenamefont [1]{#1}%
\providecommand \href@noop [0]{\@secondoftwo}%
\providecommand \href [0]{\begingroup \@sanitize@url \@href}%
\providecommand \@href[1]{\@@startlink{#1}\@@href}%
\providecommand \@@href[1]{\endgroup#1\@@endlink}%
\providecommand \@sanitize@url [0]{\catcode `\\12\catcode `\$12\catcode
  `\&12\catcode `\#12\catcode `\^12\catcode `\_12\catcode `\%12\relax}%
\providecommand \@@startlink[1]{}%
\providecommand \@@endlink[0]{}%
\providecommand \url  [0]{\begingroup\@sanitize@url \@url }%
\providecommand \@url [1]{\endgroup\@href {#1}{\urlprefix }}%
\providecommand \urlprefix  [0]{URL }%
\providecommand \Eprint [0]{\href }%
\providecommand \doibase [0]{http://dx.doi.org/}%
\providecommand \selectlanguage [0]{\@gobble}%
\providecommand \bibinfo  [0]{\@secondoftwo}%
\providecommand \bibfield  [0]{\@secondoftwo}%
\providecommand \translation [1]{[#1]}%
\providecommand \BibitemOpen [0]{}%
\providecommand \bibitemStop [0]{}%
\providecommand \bibitemNoStop [0]{.\EOS\space}%
\providecommand \EOS [0]{\spacefactor3000\relax}%
\providecommand \BibitemShut  [1]{\csname bibitem#1\endcsname}%
\let\auto@bib@innerbib\@empty
\bibitem [{\citenamefont {Szypryt}\ \emph {et~al.}(2017)\citenamefont
  {Szypryt}, \citenamefont {Meeker}, \citenamefont {Coiffard}, \citenamefont
  {Fruitwala}, \citenamefont {Bumble}, \citenamefont {Ulbricht}, \citenamefont
  {Walter}, \citenamefont {Daal}, \citenamefont {Bockstiegel}, \citenamefont
  {Collura}, \citenamefont {Zobrist}, \citenamefont {Lipartito},\ and\
  \citenamefont {Mazin}}]{paul2017}%
  \BibitemOpen
  \bibfield  {author} {\bibinfo {author} {\bibfnamefont {P.}~\bibnamefont
  {Szypryt}}, \bibinfo {author} {\bibfnamefont {S.~R.}\ \bibnamefont {Meeker}},
  \bibinfo {author} {\bibfnamefont {G.}~\bibnamefont {Coiffard}}, \bibinfo
  {author} {\bibfnamefont {N.}~\bibnamefont {Fruitwala}}, \bibinfo {author}
  {\bibfnamefont {B.}~\bibnamefont {Bumble}}, \bibinfo {author} {\bibfnamefont
  {G.}~\bibnamefont {Ulbricht}}, \bibinfo {author} {\bibfnamefont {A.~B.}\
  \bibnamefont {Walter}}, \bibinfo {author} {\bibfnamefont {M.}~\bibnamefont
  {Daal}}, \bibinfo {author} {\bibfnamefont {C.}~\bibnamefont {Bockstiegel}},
  \bibinfo {author} {\bibfnamefont {G.}~\bibnamefont {Collura}}, \bibinfo
  {author} {\bibfnamefont {N.}~\bibnamefont {Zobrist}}, \bibinfo {author}
  {\bibfnamefont {I.}~\bibnamefont {Lipartito}}, \ and\ \bibinfo {author}
  {\bibfnamefont {B.~A.}\ \bibnamefont {Mazin}},\ }\href {\doibase
  10.1364/OE.25.025894} {\bibfield  {journal} {\bibinfo  {journal} {Opt.
  Express}\ }\textbf {\bibinfo {volume} {25}},\ \bibinfo {pages} {25894}
  (\bibinfo {year} {2017})}\BibitemShut {NoStop}%
\bibitem [{\citenamefont {Strader}\ \emph {et~al.}(2016)\citenamefont
  {Strader}, \citenamefont {Archibald}, \citenamefont {Meeker}, \citenamefont
  {Szypryt}, \citenamefont {Walter}, \citenamefont {van Eyken}, \citenamefont
  {Ulbricht}, \citenamefont {Stoughton}, \citenamefont {Bumble}, \citenamefont
  {Kaplan},\ and\ \citenamefont {Mazin}}]{mattpulsar}%
  \BibitemOpen
  \bibfield  {author} {\bibinfo {author} {\bibfnamefont {M.~J.}\ \bibnamefont
  {Strader}}, \bibinfo {author} {\bibfnamefont {A.~M.}\ \bibnamefont
  {Archibald}}, \bibinfo {author} {\bibfnamefont {S.~R.}\ \bibnamefont
  {Meeker}}, \bibinfo {author} {\bibfnamefont {P.}~\bibnamefont {Szypryt}},
  \bibinfo {author} {\bibfnamefont {A.~B.}\ \bibnamefont {Walter}}, \bibinfo
  {author} {\bibfnamefont {J.~C.}\ \bibnamefont {van Eyken}}, \bibinfo {author}
  {\bibfnamefont {G.}~\bibnamefont {Ulbricht}}, \bibinfo {author}
  {\bibfnamefont {C.}~\bibnamefont {Stoughton}}, \bibinfo {author}
  {\bibfnamefont {B.}~\bibnamefont {Bumble}}, \bibinfo {author} {\bibfnamefont
  {D.~L.}\ \bibnamefont {Kaplan}}, \ and\ \bibinfo {author} {\bibfnamefont
  {B.~A.}\ \bibnamefont {Mazin}},\ }\bibfield  {title} {\enquote {\bibinfo
  {title} {{Search for optical pulsations in PSR J0337+1715}},}\ }\href
  {\doibase 10.1093/mnras/stw663} {\bibfield  {journal} {\bibinfo  {journal}
  {Monthly Notices of the Royal Astronomical Society}\ }\textbf {\bibinfo
  {volume} {459}},\ \bibinfo {pages} {427--430} (\bibinfo {year} {2016})},\
  \Eprint
  {http://arxiv.org/abs/https://academic.oup.com/mnras/article-pdf/459/1/427/8115535/stw663.pdf}
  {https://academic.oup.com/mnras/article-pdf/459/1/427/8115535/stw663.pdf}
  \BibitemShut {NoStop}%
\bibitem [{\citenamefont {Szypryt}\ \emph {et~al.}(2014)\citenamefont
  {Szypryt}, \citenamefont {Duggan}, \citenamefont {Mazin}, \citenamefont
  {Meeker}, \citenamefont {Strader}, \citenamefont {van Eyken}, \citenamefont
  {Marsden}, \citenamefont {O'Brien}, \citenamefont {Walter}, \citenamefont
  {Ulbricht}, \citenamefont {Prince}, \citenamefont {Stoughton},\ and\
  \citenamefont {Bumble}}]{paulbinary}%
  \BibitemOpen
  \bibfield  {author} {\bibinfo {author} {\bibfnamefont {P.}~\bibnamefont
  {Szypryt}}, \bibinfo {author} {\bibfnamefont {G.~E.}\ \bibnamefont {Duggan}},
  \bibinfo {author} {\bibfnamefont {B.~A.}\ \bibnamefont {Mazin}}, \bibinfo
  {author} {\bibfnamefont {S.~R.}\ \bibnamefont {Meeker}}, \bibinfo {author}
  {\bibfnamefont {M.~J.}\ \bibnamefont {Strader}}, \bibinfo {author}
  {\bibfnamefont {J.~C.}\ \bibnamefont {van Eyken}}, \bibinfo {author}
  {\bibfnamefont {D.}~\bibnamefont {Marsden}}, \bibinfo {author} {\bibfnamefont
  {K.}~\bibnamefont {O'Brien}}, \bibinfo {author} {\bibfnamefont {A.~B.}\
  \bibnamefont {Walter}}, \bibinfo {author} {\bibfnamefont {G.}~\bibnamefont
  {Ulbricht}}, \bibinfo {author} {\bibfnamefont {T.~A.}\ \bibnamefont
  {Prince}}, \bibinfo {author} {\bibfnamefont {C.}~\bibnamefont {Stoughton}}, \
  and\ \bibinfo {author} {\bibfnamefont {B.}~\bibnamefont {Bumble}},\
  }\bibfield  {title} {\enquote {\bibinfo {title} {{Direct detection of SDSS
  J0926+3624 orbital expansion with ARCONS}},}\ }\href {\doibase
  10.1093/mnras/stu137} {\bibfield  {journal} {\bibinfo  {journal} {Monthly
  Notices of the Royal Astronomical Society}\ }\textbf {\bibinfo {volume}
  {439}},\ \bibinfo {pages} {2765--2770} (\bibinfo {year} {2014})},\ \Eprint
  {http://arxiv.org/abs/https://academic.oup.com/mnras/article-pdf/439/3/2765/3850573/stu137.pdf}
  {https://academic.oup.com/mnras/article-pdf/439/3/2765/3850573/stu137.pdf}
  \BibitemShut {NoStop}%
\bibitem [{\citenamefont {Mazin}\ \emph {et~al.}(2013)\citenamefont {Mazin},
  \citenamefont {Meeker}, \citenamefont {Strader}, \citenamefont {Szypryt},
  \citenamefont {Marsden}, \citenamefont {van Eyken}, \citenamefont {Duggan},
  \citenamefont {Walter}, \citenamefont {Ulbricht}, \citenamefont {Johnson},
  \citenamefont {Bumble}, \citenamefont {O'Brien},\ and\ \citenamefont
  {Stoughton}}]{arcons}%
  \BibitemOpen
  \bibfield  {author} {\bibinfo {author} {\bibfnamefont {B.~A.}\ \bibnamefont
  {Mazin}}, \bibinfo {author} {\bibfnamefont {S.~R.}\ \bibnamefont {Meeker}},
  \bibinfo {author} {\bibfnamefont {M.~J.}\ \bibnamefont {Strader}}, \bibinfo
  {author} {\bibfnamefont {P.}~\bibnamefont {Szypryt}}, \bibinfo {author}
  {\bibfnamefont {D.}~\bibnamefont {Marsden}}, \bibinfo {author} {\bibfnamefont
  {J.~C.}\ \bibnamefont {van Eyken}}, \bibinfo {author} {\bibfnamefont {G.~E.}\
  \bibnamefont {Duggan}}, \bibinfo {author} {\bibfnamefont {A.~B.}\
  \bibnamefont {Walter}}, \bibinfo {author} {\bibfnamefont {G.}~\bibnamefont
  {Ulbricht}}, \bibinfo {author} {\bibfnamefont {M.}~\bibnamefont {Johnson}},
  \bibinfo {author} {\bibfnamefont {B.}~\bibnamefont {Bumble}}, \bibinfo
  {author} {\bibfnamefont {K.}~\bibnamefont {O'Brien}}, \ and\ \bibinfo
  {author} {\bibfnamefont {C.}~\bibnamefont {Stoughton}},\ }\href {\doibase
  10.1086/674013} {\bibfield  {journal} {\bibinfo  {journal} {Publ. Astron.
  Soc. Pac.}\ }\textbf {\bibinfo {volume} {125}},\ \bibinfo {pages}
  {1348--1361} (\bibinfo {year} {2013})}\BibitemShut {NoStop}%
\bibitem [{\citenamefont {Meeker}\ \emph {et~al.}(2018)\citenamefont {Meeker},
  \citenamefont {Mazin}, \citenamefont {Walter}, \citenamefont {Strader},
  \citenamefont {Fruitwala}, \citenamefont {Bockstiegel}, \citenamefont
  {Szypryt}, \citenamefont {Ulbricht}, \citenamefont {Coiffard}, \citenamefont
  {Bumble}, \citenamefont {Cancelo}, \citenamefont {Zmuda}, \citenamefont
  {Treptow}, \citenamefont {Wilcer}, \citenamefont {Collura}, \citenamefont
  {Dodkins}, \citenamefont {Lipartito}, \citenamefont {Zobrist}, \citenamefont
  {Bottom}, \citenamefont {Shelton}, \citenamefont {Mawet}, \citenamefont {van
  Eyken}, \citenamefont {Vasisht},\ and\ \citenamefont {Serabyn}}]{darkness}%
  \BibitemOpen
  \bibfield  {author} {\bibinfo {author} {\bibfnamefont {S.~R.}\ \bibnamefont
  {Meeker}}, \bibinfo {author} {\bibfnamefont {B.~A.}\ \bibnamefont {Mazin}},
  \bibinfo {author} {\bibfnamefont {A.~B.}\ \bibnamefont {Walter}}, \bibinfo
  {author} {\bibfnamefont {P.}~\bibnamefont {Strader}}, \bibinfo {author}
  {\bibfnamefont {N.}~\bibnamefont {Fruitwala}}, \bibinfo {author}
  {\bibfnamefont {C.}~\bibnamefont {Bockstiegel}}, \bibinfo {author}
  {\bibfnamefont {P.}~\bibnamefont {Szypryt}}, \bibinfo {author} {\bibfnamefont
  {G.}~\bibnamefont {Ulbricht}}, \bibinfo {author} {\bibfnamefont
  {G.}~\bibnamefont {Coiffard}}, \bibinfo {author} {\bibfnamefont
  {B.}~\bibnamefont {Bumble}}, \bibinfo {author} {\bibfnamefont
  {G.}~\bibnamefont {Cancelo}}, \bibinfo {author} {\bibfnamefont
  {T.}~\bibnamefont {Zmuda}}, \bibinfo {author} {\bibfnamefont
  {K.}~\bibnamefont {Treptow}}, \bibinfo {author} {\bibfnamefont
  {N.}~\bibnamefont {Wilcer}}, \bibinfo {author} {\bibfnamefont
  {G.}~\bibnamefont {Collura}}, \bibinfo {author} {\bibfnamefont
  {R.}~\bibnamefont {Dodkins}}, \bibinfo {author} {\bibfnamefont
  {I.}~\bibnamefont {Lipartito}}, \bibinfo {author} {\bibfnamefont
  {N.}~\bibnamefont {Zobrist}}, \bibinfo {author} {\bibfnamefont
  {M.}~\bibnamefont {Bottom}}, \bibinfo {author} {\bibfnamefont {J.~C.}\
  \bibnamefont {Shelton}}, \bibinfo {author} {\bibfnamefont {D.}~\bibnamefont
  {Mawet}}, \bibinfo {author} {\bibfnamefont {J.~C.}\ \bibnamefont {van
  Eyken}}, \bibinfo {author} {\bibfnamefont {G.}~\bibnamefont {Vasisht}}, \
  and\ \bibinfo {author} {\bibfnamefont {E.}~\bibnamefont {Serabyn}},\ }\href
  {\doibase 10.1088/1538-3873/aab5e7} {\bibfield  {journal} {\bibinfo
  {journal} {Publ. Astron. Soc. Pac.}\ }\textbf {\bibinfo {volume} {130}},\
  \bibinfo {pages} {065001} (\bibinfo {year} {2018})}\BibitemShut {NoStop}%
\bibitem [{\citenamefont {Walter}\ \emph {et~al.}(pted)\citenamefont {Walter},
  \citenamefont {Fruitwala}, \citenamefont {Steiger}, \citenamefont {Bailey},
  \citenamefont {Zobrist}, \citenamefont {Swimmer}, \citenamefont {Lipartito},
  \citenamefont {Smith}, \citenamefont {Meeker}, \citenamefont {Bockstiegel},
  \citenamefont {Coiffard}, \citenamefont {Dodkins}, \citenamefont {Szypryt},
  \citenamefont {Davis}, \citenamefont {Daal}, \citenamefont {Bumble},
  \citenamefont {Collura}, \citenamefont {Guyon}, \citenamefont {Lozi},
  \citenamefont {Vievard}, \citenamefont {Jovanovic}, \citenamefont
  {Martinache}, \citenamefont {Currie},\ and\ \citenamefont {Mazin}}]{mec}%
  \BibitemOpen
  \bibfield  {author} {\bibinfo {author} {\bibfnamefont {A.}~\bibnamefont
  {Walter}}, \bibinfo {author} {\bibfnamefont {N.}~\bibnamefont {Fruitwala}},
  \bibinfo {author} {\bibfnamefont {S.}~\bibnamefont {Steiger}}, \bibinfo
  {author} {\bibfnamefont {J.~I.}\ \bibnamefont {Bailey}}, \bibinfo {author}
  {\bibfnamefont {N.}~\bibnamefont {Zobrist}}, \bibinfo {author} {\bibfnamefont
  {N.}~\bibnamefont {Swimmer}}, \bibinfo {author} {\bibfnamefont
  {I.}~\bibnamefont {Lipartito}}, \bibinfo {author} {\bibfnamefont {J.~P.}\
  \bibnamefont {Smith}}, \bibinfo {author} {\bibfnamefont {S.~R.}\ \bibnamefont
  {Meeker}}, \bibinfo {author} {\bibfnamefont {C.}~\bibnamefont {Bockstiegel}},
  \bibinfo {author} {\bibfnamefont {G.}~\bibnamefont {Coiffard}}, \bibinfo
  {author} {\bibfnamefont {R.}~\bibnamefont {Dodkins}}, \bibinfo {author}
  {\bibfnamefont {P.}~\bibnamefont {Szypryt}}, \bibinfo {author} {\bibfnamefont
  {K.~K.}\ \bibnamefont {Davis}}, \bibinfo {author} {\bibfnamefont
  {M.}~\bibnamefont {Daal}}, \bibinfo {author} {\bibfnamefont {B.}~\bibnamefont
  {Bumble}}, \bibinfo {author} {\bibfnamefont {G.}~\bibnamefont {Collura}},
  \bibinfo {author} {\bibfnamefont {O.}~\bibnamefont {Guyon}}, \bibinfo
  {author} {\bibfnamefont {J.}~\bibnamefont {Lozi}}, \bibinfo {author}
  {\bibfnamefont {S.}~\bibnamefont {Vievard}}, \bibinfo {author} {\bibfnamefont
  {N.}~\bibnamefont {Jovanovic}}, \bibinfo {author} {\bibfnamefont
  {F.}~\bibnamefont {Martinache}}, \bibinfo {author} {\bibfnamefont
  {T.}~\bibnamefont {Currie}}, \ and\ \bibinfo {author} {\bibfnamefont
  {B.}~\bibnamefont {Mazin}},\ }\bibfield  {title} {\enquote {\bibinfo {title}
  {{The MKID Exoplanet Camera for Subaru SCExAO}},}\ }\href@noop {} {\bibfield
  {journal} {\bibinfo  {journal} {Publications of the Astronomical Society of
  the Pacific}\ } (\bibinfo {year} {Accepted})}\BibitemShut {NoStop}%
\bibitem [{\citenamefont {Mazin}\ \emph {et~al.}(2002)\citenamefont {Mazin},
  \citenamefont {Day}, \citenamefont {Zmuidzinas},\ and\ \citenamefont
  {LeDuc}}]{benmkidog}%
  \BibitemOpen
  \bibfield  {author} {\bibinfo {author} {\bibfnamefont {B.}~\bibnamefont
  {Mazin}}, \bibinfo {author} {\bibfnamefont {P.}~\bibnamefont {Day}}, \bibinfo
  {author} {\bibfnamefont {J.}~\bibnamefont {Zmuidzinas}}, \ and\ \bibinfo
  {author} {\bibfnamefont {H.}~\bibnamefont {LeDuc}},\ }\bibfield  {title}
  {\enquote {\bibinfo {title} {Multiplexable kinetic inductance detectors},}\
  }in\ \href@noop {} {\emph {\bibinfo {booktitle} {AIP Conference
  Proceedings}}},\ Vol.\ \bibinfo {volume} {605}\ (\bibinfo {organization}
  {American Institute of Physics},\ \bibinfo {year} {2002})\ pp.\ \bibinfo
  {pages} {309--312}\BibitemShut {NoStop}%
\bibitem [{\citenamefont {McHugh}\ \emph {et~al.}(2012)\citenamefont {McHugh},
  \citenamefont {Mazin}, \citenamefont {Serfass}, \citenamefont {Meeker},
  \citenamefont {O’Brien}, \citenamefont {Duan}, \citenamefont {Raffanti},\
  and\ \citenamefont {Werthimer}}]{seangen1}%
  \BibitemOpen
  \bibfield  {author} {\bibinfo {author} {\bibfnamefont {S.}~\bibnamefont
  {McHugh}}, \bibinfo {author} {\bibfnamefont {B.~A.}\ \bibnamefont {Mazin}},
  \bibinfo {author} {\bibfnamefont {B.}~\bibnamefont {Serfass}}, \bibinfo
  {author} {\bibfnamefont {S.}~\bibnamefont {Meeker}}, \bibinfo {author}
  {\bibfnamefont {K.}~\bibnamefont {O’Brien}}, \bibinfo {author}
  {\bibfnamefont {R.}~\bibnamefont {Duan}}, \bibinfo {author} {\bibfnamefont
  {R.}~\bibnamefont {Raffanti}}, \ and\ \bibinfo {author} {\bibfnamefont
  {D.}~\bibnamefont {Werthimer}},\ }\href {\doibase 10.1063/1.3700812}
  {\bibfield  {journal} {\bibinfo  {journal} {Review of Scientific
  Instruments}\ }\textbf {\bibinfo {volume} {83}},\ \bibinfo {pages} {044702}
  (\bibinfo {year} {2012})}\BibitemShut {NoStop}%
\bibitem [{\citenamefont {Day}\ \emph {et~al.}(2003)\citenamefont {Day},
  \citenamefont {Leduc}, \citenamefont {Mazin}, \citenamefont {Vayonakis},\
  and\ \citenamefont {Zmuidzinas}}]{day03}%
  \BibitemOpen
  \bibfield  {author} {\bibinfo {author} {\bibfnamefont {P.}~\bibnamefont
  {Day}}, \bibinfo {author} {\bibfnamefont {H.}~\bibnamefont {Leduc}}, \bibinfo
  {author} {\bibfnamefont {B.}~\bibnamefont {Mazin}}, \bibinfo {author}
  {\bibfnamefont {A.}~\bibnamefont {Vayonakis}}, \ and\ \bibinfo {author}
  {\bibfnamefont {J.}~\bibnamefont {Zmuidzinas}},\ }\bibfield  {title}
  {\enquote {\bibinfo {title} {A broadband superconducting detector suitable
  for use in large arrays},}\ }\href {\doibase 10.1038/nature02037} {\bibfield
  {journal} {\bibinfo  {journal} {Nature}\ }\textbf {\bibinfo {volume} {425}},\
  \bibinfo {pages} {817--21} (\bibinfo {year} {2003})}\BibitemShut {NoStop}%
\bibitem [{\citenamefont {Werthimer}(2011)}]{casper2011}%
  \BibitemOpen
  \bibfield  {author} {\bibinfo {author} {\bibfnamefont {D.}~\bibnamefont
  {Werthimer}},\ }\bibfield  {title} {\enquote {\bibinfo {title} {The casper
  collaboration for high-performance open source digital radio astronomy
  instrumentation},}\ }in\ \href@noop {} {\emph {\bibinfo {booktitle} {2011
  XXXth URSI general assembly and scientific symposium}}}\ (\bibinfo
  {organization} {IEEE},\ \bibinfo {year} {2011})\ pp.\ \bibinfo {pages}
  {1--4}\BibitemShut {NoStop}%
\bibitem [{\citenamefont {Hickish}\ \emph {et~al.}(2016)\citenamefont
  {Hickish}, \citenamefont {Abdurashidova}, \citenamefont {Ali}, \citenamefont
  {Buch}, \citenamefont {Chaudhari}, \citenamefont {Chen}, \citenamefont
  {Dexter}, \citenamefont {Domagalski}, \citenamefont {Ford}, \citenamefont
  {Foster} \emph {et~al.}}]{casperbig}%
  \BibitemOpen
  \bibfield  {author} {\bibinfo {author} {\bibfnamefont {J.}~\bibnamefont
  {Hickish}}, \bibinfo {author} {\bibfnamefont {Z.}~\bibnamefont
  {Abdurashidova}}, \bibinfo {author} {\bibfnamefont {Z.}~\bibnamefont {Ali}},
  \bibinfo {author} {\bibfnamefont {K.~D.}\ \bibnamefont {Buch}}, \bibinfo
  {author} {\bibfnamefont {S.~C.}\ \bibnamefont {Chaudhari}}, \bibinfo {author}
  {\bibfnamefont {H.}~\bibnamefont {Chen}}, \bibinfo {author} {\bibfnamefont
  {M.}~\bibnamefont {Dexter}}, \bibinfo {author} {\bibfnamefont {R.~S.}\
  \bibnamefont {Domagalski}}, \bibinfo {author} {\bibfnamefont
  {J.}~\bibnamefont {Ford}}, \bibinfo {author} {\bibfnamefont {G.}~\bibnamefont
  {Foster}},  \emph {et~al.},\ }\bibfield  {title} {\enquote {\bibinfo {title}
  {A decade of developing radio-astronomy instrumentation using casper
  open-source technology},}\ }\href@noop {} {\bibfield  {journal} {\bibinfo
  {journal} {Journal of Astronomical Instrumentation}\ }\textbf {\bibinfo
  {volume} {5}},\ \bibinfo {pages} {1641001} (\bibinfo {year}
  {2016})}\BibitemShut {NoStop}%
\bibitem [{Note1()}]{Note1}%
  \BibitemOpen
  \bibinfo {note} {There is a fixed 3 dB of attenuation in addition to the
  programmable attenuators, so the power at the end of the chain is given by
  $P_{out} = P_{in} + 60\ dB - A_{prog} - 3\ dB$}\BibitemShut {NoStop}%
\bibitem [{\citenamefont {{Moseley}}\ \emph {et~al.}(1988)\citenamefont
  {{Moseley}}, \citenamefont {{Kelley}}, \citenamefont {{Schoelkopf}},
  \citenamefont {{Szymkowiak}}, \citenamefont {{McCammon}},\ and\ \citenamefont
  {{Zhang}}}]{origoptfilt}%
  \BibitemOpen
  \bibfield  {author} {\bibinfo {author} {\bibfnamefont {S.~H.}\ \bibnamefont
  {{Moseley}}}, \bibinfo {author} {\bibfnamefont {R.~L.}\ \bibnamefont
  {{Kelley}}}, \bibinfo {author} {\bibfnamefont {R.~J.}\ \bibnamefont
  {{Schoelkopf}}}, \bibinfo {author} {\bibfnamefont {A.~E.}\ \bibnamefont
  {{Szymkowiak}}}, \bibinfo {author} {\bibfnamefont {D.}~\bibnamefont
  {{McCammon}}}, \ and\ \bibinfo {author} {\bibfnamefont {J.}~\bibnamefont
  {{Zhang}}},\ }\bibfield  {title} {\enquote {\bibinfo {title} {Advances toward
  high spectral resolution quantum x-ray calorimetry},}\ }\href@noop {}
  {\bibfield  {journal} {\bibinfo  {journal} {IEEE Transactions on Nuclear
  Science}\ }\textbf {\bibinfo {volume} {35}},\ \bibinfo {pages} {59--64}
  (\bibinfo {year} {1988})}\BibitemShut {NoStop}%
\bibitem [{\citenamefont {Alpert}\ \emph {et~al.}(2013)\citenamefont {Alpert},
  \citenamefont {Horansky}, \citenamefont {Bennett}, \citenamefont {Doriese},
  \citenamefont {Fowler}, \citenamefont {Hoover}, \citenamefont {Rabin},\ and\
  \citenamefont {Ullom}}]{alpert}%
  \BibitemOpen
  \bibfield  {author} {\bibinfo {author} {\bibfnamefont {B.~K.}\ \bibnamefont
  {Alpert}}, \bibinfo {author} {\bibfnamefont {R.~D.}\ \bibnamefont
  {Horansky}}, \bibinfo {author} {\bibfnamefont {D.~A.}\ \bibnamefont
  {Bennett}}, \bibinfo {author} {\bibfnamefont {W.~B.}\ \bibnamefont
  {Doriese}}, \bibinfo {author} {\bibfnamefont {J.~W.}\ \bibnamefont {Fowler}},
  \bibinfo {author} {\bibfnamefont {A.~S.}\ \bibnamefont {Hoover}}, \bibinfo
  {author} {\bibfnamefont {M.~W.}\ \bibnamefont {Rabin}}, \ and\ \bibinfo
  {author} {\bibfnamefont {J.~N.}\ \bibnamefont {Ullom}},\ }\bibfield  {title}
  {\enquote {\bibinfo {title} {Note: Operation of gamma-ray microcalorimeters
  at elevated count rates using filters with constraints},}\ }\href {\doibase
  10.1063/1.4806802} {\bibfield  {journal} {\bibinfo  {journal} {Review of
  Scientific Instruments}\ }\textbf {\bibinfo {volume} {84}},\ \bibinfo {pages}
  {056107} (\bibinfo {year} {2013})},\ \Eprint
  {http://arxiv.org/abs/https://doi.org/10.1063/1.4806802}
  {https://doi.org/10.1063/1.4806802} \BibitemShut {NoStop}%
\bibitem [{\citenamefont {Zobrist}\ \emph {et~al.}(2019)\citenamefont
  {Zobrist}, \citenamefont {Eom}, \citenamefont {Day}, \citenamefont {Mazin},
  \citenamefont {Meeker}, \citenamefont {Bumble}, \citenamefont {LeDuc},
  \citenamefont {Coiffard}, \citenamefont {Szypryt}, \citenamefont {Fruitwala},
  \citenamefont {Lipartito},\ and\ \citenamefont {Bockstiegel}}]{zobrist}%
  \BibitemOpen
  \bibfield  {author} {\bibinfo {author} {\bibfnamefont {N.}~\bibnamefont
  {Zobrist}}, \bibinfo {author} {\bibfnamefont {B.~H.}\ \bibnamefont {Eom}},
  \bibinfo {author} {\bibfnamefont {P.}~\bibnamefont {Day}}, \bibinfo {author}
  {\bibfnamefont {B.~A.}\ \bibnamefont {Mazin}}, \bibinfo {author}
  {\bibfnamefont {S.~R.}\ \bibnamefont {Meeker}}, \bibinfo {author}
  {\bibfnamefont {B.}~\bibnamefont {Bumble}}, \bibinfo {author} {\bibfnamefont
  {H.~G.}\ \bibnamefont {LeDuc}}, \bibinfo {author} {\bibfnamefont
  {G.}~\bibnamefont {Coiffard}}, \bibinfo {author} {\bibfnamefont
  {P.}~\bibnamefont {Szypryt}}, \bibinfo {author} {\bibfnamefont
  {N.}~\bibnamefont {Fruitwala}}, \bibinfo {author} {\bibfnamefont
  {I.}~\bibnamefont {Lipartito}}, \ and\ \bibinfo {author} {\bibfnamefont
  {C.}~\bibnamefont {Bockstiegel}},\ }\bibfield  {title} {\enquote {\bibinfo
  {title} {Wide-band parametric amplifier readout and resolution of optical
  microwave kinetic inductance detectors},}\ }\href {\doibase
  10.1063/1.5098469} {\bibfield  {journal} {\bibinfo  {journal} {Applied
  Physics Letters}\ }\textbf {\bibinfo {volume} {115}},\ \bibinfo {pages}
  {042601} (\bibinfo {year} {2019})},\ \Eprint
  {http://arxiv.org/abs/https://doi.org/10.1063/1.5098469}
  {https://doi.org/10.1063/1.5098469} \BibitemShut {NoStop}%
\bibitem [{\citenamefont {Gao}\ \emph {et~al.}(2007)\citenamefont {Gao},
  \citenamefont {Zmuidzinas}, \citenamefont {Mazin}, \citenamefont {LeDuc},\
  and\ \citenamefont {Day}}]{gaotls1}%
  \BibitemOpen
  \bibfield  {author} {\bibinfo {author} {\bibfnamefont {J.}~\bibnamefont
  {Gao}}, \bibinfo {author} {\bibfnamefont {J.}~\bibnamefont {Zmuidzinas}},
  \bibinfo {author} {\bibfnamefont {B.~A.}\ \bibnamefont {Mazin}}, \bibinfo
  {author} {\bibfnamefont {H.~G.}\ \bibnamefont {LeDuc}}, \ and\ \bibinfo
  {author} {\bibfnamefont {P.~K.}\ \bibnamefont {Day}},\ }\bibfield  {title}
  {\enquote {\bibinfo {title} {Noise properties of superconducting coplanar
  waveguide microwave resonators},}\ }\href@noop {} {\bibfield  {journal}
  {\bibinfo  {journal} {Applied Physics Letters}\ }\textbf {\bibinfo {volume}
  {90}},\ \bibinfo {pages} {102507} (\bibinfo {year} {2007})}\BibitemShut
  {NoStop}%
\bibitem [{\citenamefont {Gao}\ \emph {et~al.}(2008)\citenamefont {Gao},
  \citenamefont {Daal}, \citenamefont {Vayonakis}, \citenamefont {Kumar},
  \citenamefont {Zmuidzinas}, \citenamefont {Sadoulet}, \citenamefont {Mazin},
  \citenamefont {Day},\ and\ \citenamefont {Leduc}}]{gaotls2}%
  \BibitemOpen
  \bibfield  {author} {\bibinfo {author} {\bibfnamefont {J.}~\bibnamefont
  {Gao}}, \bibinfo {author} {\bibfnamefont {M.}~\bibnamefont {Daal}}, \bibinfo
  {author} {\bibfnamefont {A.}~\bibnamefont {Vayonakis}}, \bibinfo {author}
  {\bibfnamefont {S.}~\bibnamefont {Kumar}}, \bibinfo {author} {\bibfnamefont
  {J.}~\bibnamefont {Zmuidzinas}}, \bibinfo {author} {\bibfnamefont
  {B.}~\bibnamefont {Sadoulet}}, \bibinfo {author} {\bibfnamefont {B.~A.}\
  \bibnamefont {Mazin}}, \bibinfo {author} {\bibfnamefont {P.~K.}\ \bibnamefont
  {Day}}, \ and\ \bibinfo {author} {\bibfnamefont {H.~G.}\ \bibnamefont
  {Leduc}},\ }\href {\doibase 10.1063/1.2906373} {\bibfield  {journal}
  {\bibinfo  {journal} {Appl. Phys. Lett.}\ }\textbf {\bibinfo {volume} {92}},\
  \bibinfo {pages} {152505} (\bibinfo {year} {2008})}\BibitemShut {NoStop}%
\bibitem [{\citenamefont {Zmuidzinas}(2012)}]{jonasreview}%
  \BibitemOpen
  \bibfield  {author} {\bibinfo {author} {\bibfnamefont {J.}~\bibnamefont
  {Zmuidzinas}},\ }\bibfield  {title} {\enquote {\bibinfo {title}
  {Superconducting microresonators: Physics and applications},}\ }\href
  {\doibase 10.1146/annurev-conmatphys-020911-125022} {\bibfield  {journal}
  {\bibinfo  {journal} {Annual Review of Condensed Matter Physics}\ }\textbf
  {\bibinfo {volume} {3}},\ \bibinfo {pages} {169--214} (\bibinfo {year}
  {2012})}\BibitemShut {NoStop}%
\bibitem [{\citenamefont {Golwala}(2000)}]{golwalathesis}%
  \BibitemOpen
  \bibfield  {author} {\bibinfo {author} {\bibfnamefont {S.~R.}\ \bibnamefont
  {Golwala}},\ }\emph {\bibinfo {title} {{Exclusion limits on the WIMP nucleon
  elastic scattering cross-section from the Cryogenic Dark Matter Search}}},\
  \href {\doibase 10.2172/1421437} {Ph.D. thesis},\ \bibinfo  {school} {UC,
  Berkeley} (\bibinfo {year} {2000})\BibitemShut {NoStop}%
\bibitem [{\citenamefont {Strader}(2016)}]{mattthesis}%
  \BibitemOpen
  \bibfield  {author} {\bibinfo {author} {\bibfnamefont {M.}~\bibnamefont
  {Strader}},\ }\emph {\bibinfo {title} {Digital Readout for Microwave Kinetic
  InductanceDetectors and Applications in High Time ResolutionAstronomy}},\
  \href@noop {} {Ph.D. thesis},\ \bibinfo  {school} {University of California
  Santa Barbara} (\bibinfo {year} {2016})\BibitemShut {NoStop}%
\end{thebibliography}%

\end{document}